\documentclass[conference]{IEEEtran}
\IEEEoverridecommandlockouts
\usepackage{cite}
\usepackage{amsmath,amssymb,amsfonts}
\usepackage{algorithmic}
\usepackage{graphicx}
\usepackage{textcomp}
\usepackage{xcolor}
\usepackage{booktabs}
\usepackage{pifont}
\usepackage{multirow}
\usepackage{multicol}
\usepackage[T1]{fontenc}
\def\BibTeX{{\rm B\kern-.05em{\sc i\kern-.025em b}\kern-.08em
    T\kern-.1667em\lower.7ex\hbox{E}\kern-.125emX}}

\usepackage{listings}
\begin{document}

\title{Speedbumps: Rejection Attacks on Speculative Decoding}

\author{
\IEEEauthorblockN{Adam Y. J. Jones\IEEEauthorrefmark{1}}
\IEEEauthorblockA{\textit{Imperial College London}\\
London, UK \\
adam.y.jones@imperial.ac.uk}
\and
\IEEEauthorblockN{Yu Yuan\IEEEauthorrefmark{1}}
\IEEEauthorblockA{\textit{Imperial College London}\\
London, UK \\
yu.yuan125@imperial.ac.uk}
\and
\IEEEauthorblockN{Sergio Maffeis}
\IEEEauthorblockA{\textit{Imperial College London}\\
London, UK \\
sergio.maffeis@imperial.ac.uk}
\thanks{\IEEEauthorrefmark{1}These authors contributed equally to this work.}
}

\maketitle

\begin{abstract}
Speculative decoding is a popular technique for increasing the speed and reducing the costs of large language model (LLM) inference by verifying multiple draft tokens in a single target-model forward pass. The resulting benefit depends on the ability of the drafter to approximate the target model's distribution. In this work, we study Speculative Rejection Attacks (SRAs), a novel class of attacks that cause draft and target models to disagree more often, resulting in fewer draft tokens being accepted per draft cycle. This leads to more target model forward passes needed per generated token, slowing down inference and increasing costs for the victim. We introduce two attacks which append an adversarial suffix to attacker-controlled content to degrade speculative decoding on a victim's prompts. Both attacks optimise the expected length of the accepted speculative prefix, estimating per-depth acceptance from the target's probability of the drafted proposals (Speedbump-P) or from the overlap between the draft and target distributions (Speedbump-D). In some cases, attacks degrade speculative decoding to the point of being slower than autoregressive decoding. The degradation reduces the output quality - regularisation restores output quality but gives up most of the degradation, trading effectiveness for stealthiness. Additionally, the suffixes remain effective under sampling, and transfer across drafters (Speedbump-P) or across target models sharing a drafter (Speedbump-D). These findings identify the draft-target interaction of speculative decoding as a realistic attack surface through which adversarial inputs can inflate inference costs.
\end{abstract}


\section{Introduction}

The rapid adoption of large language models (LLMs) has highlighted inference efficiency as a key challenge. In standard autoregressive decoding, each token is produced sequentially through a separate forward pass of the model, creating a fundamental latency bottleneck. Speculative Decoding (SD) \cite{leviathan2023fast,chen2023accelerating} addresses this by introducing a cheap proposal mechanism that generates candidate future tokens, which are subsequently verified by the target model in parallel using a single forward pass. When the proposed tokens are accepted, multiple tokens can be committed in a single verification step, reducing the number of sequential target-model executions required for generation.

The acceleration achieved by SD depends on the agreement between the proposal mechanism and the target model. High agreement enables more proposed tokens to be accepted and allows the cost of verification to be amortised over multiple generated tokens. In contrast, low agreement causes earlier rejection and diminishes the benefit of speculation.

SD is supported by widely used serving frameworks \cite{vllm,sglang} - many operators enable it as a free cost-saving mechanism as it provably leaves the output unchanged. However, the savings are not guaranteed: they depend on how many drafted tokens the target accepts. We ask whether an adversary can exploit this to slow down inference and inflate the victim's inference costs.

\begin{figure*}
    \centering
    \includegraphics[width=0.85\linewidth]{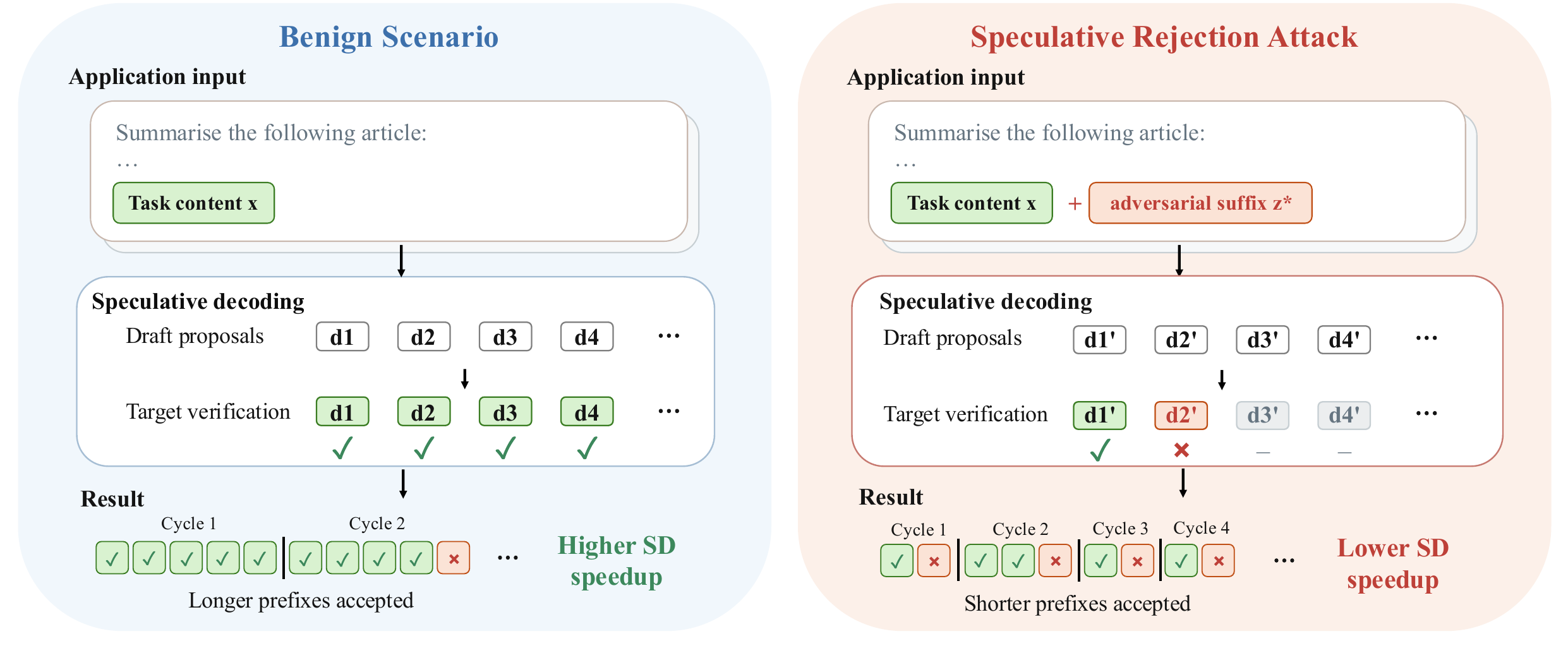}
    \caption{Illustration of the effect of a Speculative Rejection Attack.}
    \label{fig:ssas_diagram}
\end{figure*}

We study \emph{Speculative Rejection Attacks} (SRAs), in which an attacker who controls content processed by an LLM application, such as a document or webpage, aims to raise the victim's cost of generating each output token (Figure \ref{fig:ssas_diagram}. We measure this cost as the target-model forward passes and GPU time spent per generated token. Each rejected draft forces additional target passes, and the victim pays for the resulting GPU time as energy, as rental charges, or as serving capacity lost to other users. Since the attacker's goal is cost, they need not preserve output quality. Quality matters only when a visibly altered response could make the attack easier to detect.

Prior attacks may raise cost by lengthening the output \cite{engorgio,nonhalting,loopllm,overthink}. SRAs are orthogonal: length attacks increase the number of tokens, whereas SRAs increase the cost of each token, so the two may compose multiplicatively. SRAs also remain effective when a response already reaches the generation cap, where a length attack has nothing left to gain.

We introduce \emph{Speedbump}, an adversarial suffix attack whose objective is the expected number of draft tokens accepted per cycle. The objective reflects the prefix structure of verification, in which a draft token saves a target pass only if all preceding draft tokens are accepted. We instantiate it with two differentiable proxies: \emph{Speedbump-P}, which uses only the drafted tokens, and \emph{Speedbump-D}, which also uses the drafter's distribution. A single suffix is optimised per task and reused as is on unseen inputs, leaving the victim's model, prompt, and decoder unchanged.

Across seven SD methods spanning independent drafters, prediction heads, feature-level drafters, and model-free retrieval, across three tasks, Speedbump-P reduces speed-up by 12.2--36.8\% under greedy decoding, depending on the task. In the strongest case, the target model needs 3.2 times as many forward passes per generated token (EAGLE-3 on summarisation). Under sampling, some attacks make SD slower than autoregressive decoding, so that enabling SD costs the victim more than disabling it. Preserving output quality costs the attacker most of the slowdown, and the suffixes transfer partially to unseen drafters and target models.

Our contributions are as follows:
\begin{itemize}
    \item We characterise Speculative Rejection Attacks (SRAs), which exploit speculative decoding to raise the per-token inference cost.

    \item We introduce \emph{Speedbump}, a prefix-aware adversarial attack that reduces draft token acceptance. We instantiate this objective with two differentiable proxies requiring different levels of access to the drafter.

    \item We provide a broad empirical study of this attack surface across different configurations and threat models.

    \item We characterise the stealthiness of the attacks, as well as proposing potential countermeasures and mitigations once such an attack is detected.

\end{itemize}

\section{Related Work}

\subsection{Speculative Decoding}
Research on speculative decoding has explored a range of proposal mechanisms for generating candidate continuations. Early methods use a separate lightweight language model as an independent drafter \cite{leviathan2023fast,chen2023accelerating}, with subsequent work improving the efficiency and alignment of such drafters through specialised training and adaptation \cite{distillspec,onlineSD}.

A broad class of later methods integrates drafting more closely with the target model. Medusa \cite{medusa} and Hydra \cite{hydra} attach lightweight prediction heads to target-model representations, while the EAGLE family \cite{eagle,eagle2,eagle3} constructs lightweight drafters from internal target-model features. Other approaches reuse the target model itself for drafting through early exiting \cite{draftverify} or layer skipping \cite{layerskip,kangaroo}, generating proposals from shallow or reduced execution paths before verification with the full model. Beyond model-based drafting, some approaches construct speculative proposals by retrieving existing text, drawing from the current context \cite{llma,pld}, task-relevant references, or external retrieval sources \cite{rest}.

Orthogonal to the proposal mechanism, speculative decoding methods also vary in how candidates are organised and verified. Proposals may form a single chain or a branching tree, with tree-structured methods verifying multiple candidate continuations within a single target-model pass. Verification also depends on the decoding regime: under greedy decoding, proposed tokens are retained while they agree with the target model's greedy continuation, whereas sampling-based methods use probabilistic acceptance and correction rules to preserve the target distribution. These choices introduce different trade-offs between candidate coverage, proposal cost, acceptance, and verification overhead.

\subsection{Exploiting Speculative Decoding}
The accept-reject behaviour of speculative decoding can create observable side channels. Carlini and Nasr \cite{carlini2024remote} show that efficient inference mechanisms, including speculative decoding, can leak information about user queries through remotely observable timing differences. Wei et al. \cite{wei2026speculation} further demonstrate that speculation patterns can be exploited to fingerprint queries and, in retrieval-based settings, reveal information from the proposal datastore. Together, these works show that speculative decoding can introduce security risks beyond its intended efficiency benefits.

\subsection{Latency Attacks}
Latency attacks seek to increase the computational cost of model inference, reducing serving throughput and potentially degrading service availability. This connects to algorithmic-complexity denial-of-service attacks \cite{crosby2003dos}, where adversarially chosen inputs trigger disproportionately expensive execution. In machine learning, prior work has explored several ways of forcing models towards more expensive execution: DeepSloth \cite{deepsloth} delays early-exit decisions in adaptive multi-exit networks, while sponge examples \cite{shumailov2021sponge} more broadly induce high-latency and high-energy execution.

For autoregressive LLMs, prior attacks \cite{engorgio,nonhalting,loopllm} induce longer generations, for example by preventing termination or causing repetitive and non-halting behaviour. Attacks on reasoning models \cite{overthink,excessivereasoning} similarly induce substantially longer reasoning traces, in some cases while preserving useful model outputs.

Speculative decoding introduces a distinct attack surface for latency degradation. We study this setting using a new expected-prefix approach to adversarially reduce speculative-decoding efficiency. Concurrent preprints Mistletoe~\cite{sun2026mistletoestealthyaccelerationcollapseattacks} and ADSD~\cite{ADSDwang2026adversarialpromptsacceptancecollapse} independently consider a similar threat, but differ in attack objective and threat model. We compare directly against their alternative objectives in Section~\ref{sec:objective_ablations}.

\section{Preliminaries}
\label{sec:preliminaries}

Speculative decoding proceeds through repeated draft-and-verify cycles. Each cycle starts from the current committed prefix: the proposal mechanism generates one or more candidate continuations, the target model verifies these candidates in parallel, and the accepted tokens are committed before generation proceeds to the next cycle. Let $p_t(\cdot)$ denote the target model's next-token distribution at a speculative context $t$, and let $G_t$ denote the set of candidate child tokens exposed for verification at that context. For a chain, $G_t$ contains a single proposed token, whereas a tree may expose multiple sibling candidates under the same context. When the proposal mechanism exposes a next-token distribution, we additionally denote it by $q_t(\cdot)$.

\paragraph{Sampling-based verification}
\label{sec:sampling_based_verification}
In standard sampling-based speculative decoding, a proposal $v \sim q_t$ is accepted with probability
\begin{equation}
    a_t(v)=\min\left(1,\frac{p_t(v)}{q_t(v)}\right).
\end{equation}
If the proposal is rejected, a replacement token is sampled from the residual distribution
\begin{equation}
    r_t(u)=\frac{[p_t(u)-q_t(u)]_+}{\sum_w [p_t(w)-q_t(w)]_+},
\end{equation}
which ensures that the resulting token follows the target distribution \cite{leviathan2023fast,chen2023accelerating}. Averaging the acceptance probability over proposals from $q_t$ gives
\begin{equation}
\mathbb{E}_{v\sim q_t}[a_t(v)]=\sum_v\min\{p_t(v),q_t(v)\}=1-\mathrm{TV}(p_t,q_t).
\end{equation}
where $\mathrm{TV}$ is the total variation between the distributions $p_t$ and $q_t$. Hence, for a fixed speculative context, the expected acceptance probability is exactly the overlap between the draft and target distributions. The equations above describe the standard single-proposal acceptance rule. Tree-structured sampling methods extend this setting to multiple candidates, with the exact acceptance and correction procedure depending on how the proposal tree is constructed, while still preserving the target distribution.

\paragraph{Greedy verification}
Under greedy decoding, let
\begin{equation}
    y_t=\arg\max_v p_t(v)
\end{equation}
denote the target model's greedy next-token prediction. Verification succeeds at context $t$ when the target's greedy token is present among the exposed proposals,
\begin{equation}
    I_t^{\mathrm{greedy}}=\mathbf{1}\{y_t\in G_t\}.
\end{equation}
For a chain with a single greedy proposal $\hat{v}_t$, this reduces to the condition $\hat{v}_t=y_t$. For a tree, if a child labelled with $y_t$ exists at the current node, that child is accepted and verification proceeds from the corresponding child context; otherwise, speculative acceptance terminates at that node. Although the tree may contain multiple candidate branches, greedy verification therefore commits at most one path through the tree.

\paragraph{Accepted prefixes}
Acceptance within a speculative cycle is prefix-structured. In a chain, a proposal at speculative depth $i$ can contribute to the output only if all preceding proposals in that cycle have also been accepted. Tree-based proposals permit several candidate branches, but committed progress similarly follows a single accepted path through the proposal tree.

Let $L_c$ denote the number of speculative tokens committed in cycle $c$. The total number of tokens committed by the cycle is
\begin{equation}
    A_c=L_c+1,
\end{equation}
where the additional token is obtained from the target-model forward pass used for verification: it is either a fallback token following rejection or, when all proposals are accepted, a bonus token produced from the final target prediction. Thus, each cycle commits at least one token without requiring an additional target-model forward pass. We define mean accepted length (MAL) across $C$ speculative cycles as
\begin{equation}
\label{eq:MAL}
    \mathrm{MAL}=\frac{1}{C}\sum_{c=1}^{C}A_c.
\end{equation}
Higher MAL therefore corresponds to more generation progress per target-model verification. Since each cycle commits at least one token, $\mathrm{MAL}\geq 1$, with repeated early rejection driving MAL towards this minimum.

\section{Methodology}
\begin{figure*}
    \centering
    \includegraphics[width=0.85\linewidth]{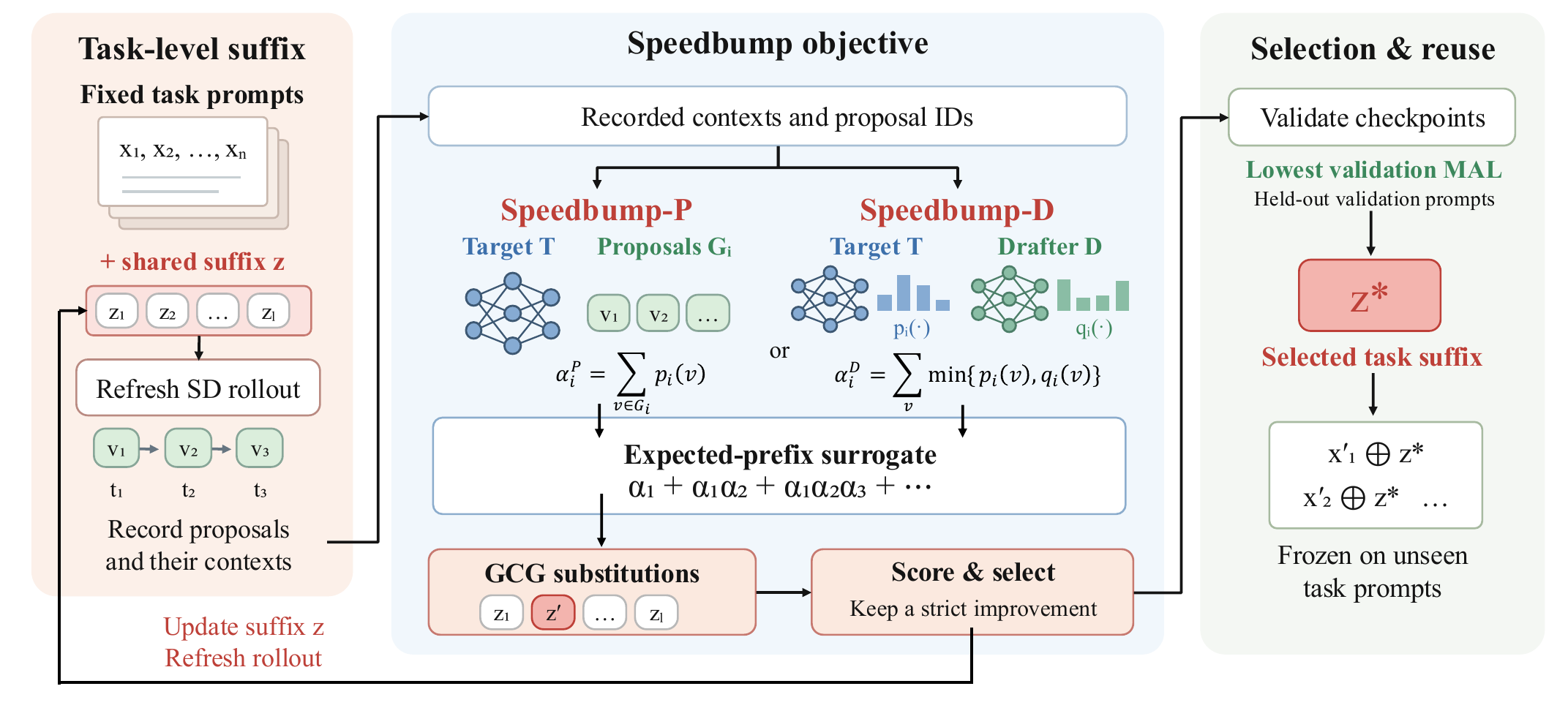}
    \caption{Overview of the Speedbump optimisation pipeline.}
    \label{fig:method_diagram}
\end{figure*}

Speedbump attacks insert an optimised suffix into attacker-controlled task content to reduce the number of draft tokens accepted by the victim's speculative decoder. We first define the threat model (Section~\ref{sec:threat_model}) and derive an expected-length attack objective (Section~\ref{sec:speedbump}), then instantiate it using two acceptance proxies with different drafter-access requirements (Section~\ref{sec:acceptance_proxies}). Finally, we describe the suffix search procedure (Section~\ref{sec:suffix_search}) and an optional target-distribution regulariser (Section~\ref{sec:target_regularisation}). Figure~\ref{fig:method_diagram} provides an overview of the attack optimisation process.

\subsection{Threat Model}
\label{sec:threat_model}
We consider an attacker who controls content that an LLM application inserts into an otherwise fixed prompt. Let $x$ denote task content supplied to an application-specific prompt template, or \emph{harness}. The attacker appends a token sequence $z$ to this content, producing $x^{z} = x \oplus z$, while leaving the surrounding instructions, model configuration, and speculative decoding system unchanged. The suffix remains inside the application-provided content field: the attacker does not modify the system prompt, the harness instructions, or an assistant prefill. Appendix~\ref{appendix:prompts} shows the suffix placement for each task.

This setting models applications that process content originating from partially or fully untrusted sources. For example, a news publisher could place an adversarial suffix inside an article that is later processed by an LLM-based summarisation service, with the goal of increasing the computational cost of processing that content. Similar settings arise when LLM applications consume externally supplied documents, webpages, or retrieved text.

For each task, the attacker optimises a single \emph{task-level} suffix offline over a fixed set of optimisation prompts. The resulting suffix is then frozen and applied unchanged to unseen inputs of the same task. During optimisation, the attacker has white-box access to the target model, and knows the task and the speculative decoder in use. Speedbump-P additionally requires the drafter's proposed tokens and their contexts, and Speedbump-D the drafter's full next-token distributions (Section~\ref{sec:acceptance_proxies}).

Our threat model differs from prior work in two ways. Firstly, Mistletoe \cite{sun2026mistletoestealthyaccelerationcollapseattacks} optimises a separate suffix for each input of a given task, whereas our suffix is optimised per task and shared between inputs. Secondly, Mistletoe appends its optimised suffix directly to the clean input, while the prompt template reported by ADSD \cite{ADSDwang2026adversarialpromptsacceptancecollapse} places the adversarial content after the assistant turn marker. By contrast, our threat model keeps the suffix within the task content consumed by the application harness. This gives the attacker less direct control over the generation boundary, and more closely reflects an attacker who controls external content rather than the LLM interface itself.

\subsection{Expected-Length Objective}
\label{sec:speedbump}
The attacker's aim is to reduce the number of draft tokens accepted per cycle. We use the prefix structure of speculative verification to construct a cycle-level attack objective. This objective is shared by both of our attacks, which differ only in how acceptance is estimated at each depth (Section~\ref{sec:acceptance_proxies}).

Consider a speculative cycle $c$ with $m_c$ scored draft depths, and let $L_c$ denote the number of accepted draft tokens in that cycle. Verification commits a single path through the draft (a chain being the special case of a tree with one proposal per depth), and we index depths along this path. Suppose $r_{c,i}$ is the conditional probability of accepting the proposal at depth $i$, given that the preceding draft prefix has been accepted. Then
\begin{equation}
\Pr(L_c \geq i)
=
\prod_{j=1}^{i} r_{c,j},
\end{equation}
and therefore
\begin{equation}
\mathbb{E}[L_c]
=
\sum_{i=1}^{m_c}
\Pr(L_c \geq i)
=
\sum_{i=1}^{m_c}
\prod_{j=1}^{i} r_{c,j}.
\label{eq:expected_prefix_length}
\end{equation}
This expression is independent of the verification rule, which only determines $r_{c,i}$. It captures prefix reachability: reducing acceptance at an early depth lowers the probability of reaching every subsequent depth. The target-generated fallback or bonus token included in MAL is excluded.

The conditional acceptance probabilities $r_{c,i}$ cannot in general be optimised directly. Their form depends on the verification rule, they may require unavailable drafter information, and under deterministic rules they are indicators with no useful gradient with respect to the suffix. To obtain a differentiable attack objective, we replace them with local acceptance proxies $\alpha_{c,i}\in[0,1]$. For a prompt $x$ whose rollout contains $C_x$ speculative cycles, we define
\begin{equation}
J_{\mathrm{Speedbump}}(x)
=
-\frac{1}{C_x}
\sum_{c=1}^{C_x}
\sum_{i=1}^{m_c}
\prod_{j=1}^{i}
\alpha_{c,j}.
\label{eq:speedbump_objective}
\end{equation}

The attack maximises $J_{\mathrm{Speedbump}}$, averaged over the optimisation prompts. In each cycle, we score only the $m_c$ depths actually reached in the rollout: the accepted draft tokens and the first rejected proposal. Proposals conditioned on an unreached prefix do not contribute.

The attack is therefore determined by the choice of $\alpha_{c,i}$. A suitable proxy must be low where acceptance is unlikely, differentiable with respect to the suffix, and computable from the information the attacker has about the drafter.

\subsection{Acceptance Proxies}
\label{sec:acceptance_proxies}

We construct two local acceptance proxies for Equation~\ref{eq:speedbump_objective}, motivated by the verification rules in Section~\ref{sec:preliminaries} and defining our two attacks. The proposal mass proxy, used in our attack Speedbump-P, relaxes the acceptance event of greedy verification. The draft-target overlap proxy, used in our attack Speedbump-D, is the acceptance probability of sampling-based verification. Both are applicable under any verification rule, but differ in the information they require about the drafter. Let $t_{c,i}$ denote the scored context at depth $i$ of cycle $c$, and let $G_{c,i}$ denote the set of distinct tokens proposed for verification at that context.

\subsubsection{Proposal Mass Proxy (Speedbump-P)}
\label{sec:speedbump_p}

Our proposal-aware objective uses the proposals exposed by the speculative decoder. We define the local acceptance proxy as the relaxation of the greedy verification indicator to the target probability mass assigned to the proposed set:
\begin{equation}
    \alpha^{P}_{c,i}
    =
    \sum_{v \in G_{c,i}}
    p_{t_{c,i}}(v).
    \label{eq:speedbump_p_proxy}
\end{equation}

For a chain-based decoder, $G_{c,i}$ contains a single proposal $v_{c,i}$, giving $\alpha^{P}_{c,i}=p_{t_{c,i}}(v_{c,i})$; for tree-based decoders, sibling proposals sharing the same context are aggregated by their total target probability mass. The proxy therefore measures the target support assigned to the proposals presented for verification. Substituting $\alpha^{P}_{c,i}$ into Equation~\ref{eq:speedbump_objective} defines Speedbump-P.

Speedbump-P requires only the proposed token identities and their conditioning contexts, together with gradients through the target model. It does not require the full drafter distribution or a differentiable proposal mechanism, and therefore also applies to model-free proposal mechanisms such as copying or retrieval when their proposals are observable.

\subsubsection{Draft-Target Overlap Proxy (Speedbump-D)}
\label{sec:speedbump_d}

Our distribution-aware objective uses the full target and drafter distributions at each scored context. Let $p_{t_{c,i}}(\cdot)$ and $q_{t_{c,i}}(\cdot)$ denote their next-token distributions at depth $i$ of cycle $c$. Under sampling-based verification with a single proposal drawn from $q$, the expected acceptance probability is their distributional overlap (Section~\ref{sec:sampling_based_verification}), which we take as the local acceptance proxy:
\begin{equation}
    \alpha^{D}_{c,i}
    =
    \sum_v
    \min\left\{
        p_{t_{c,i}}(v),
        q_{t_{c,i}}(v)
    \right\}
    =
    1-\mathrm{TV}
    \left(
        p_{t_{c,i}},
        q_{t_{c,i}}
    \right).
    \label{eq:speedbump_d_proxy}
\end{equation}
Under sampling-based verification, this proxy is exact. Under other verification rules such as greedy verification or tree-structured sampling with several proposals per depth, it instead provides a smooth, differentiable measure of draft-target agreement over the complete distributions. Substituting $\alpha^{D}_{c,i}$ into Equation~\ref{eq:speedbump_objective} defines Speedbump-D.

Speedbump-D requires access to the drafter's full next-token distribution and gradients through both models, and therefore does not apply to model-free drafters. It may also miss cases where the distributions differ overall but agree on the tokens actually tested by verification.

\subsection{Suffix Search}
\label{sec:suffix_search}

Each optimisation run targets a single task, decoder, and attack configuration, and returns a suffix $z=(z_1,\ldots,z_L)$ of length $L$. We optimise over a set $\mathcal{D}_{\mathrm{opt}}$ containing $n_{\mathrm{opt}}$ optimisation prompts and select the final suffix using a disjoint validation set $\mathcal{D}_{\mathrm{val}}$ containing $n_{\mathrm{val}}$ prompts. Using gradient-guided coordinate search following GCG~\cite{zou2023universal}, we maximise the objective
\begin{equation}
J(z)=\frac{1}{n_{\mathrm{opt}}}\sum_{x\in\mathcal{D}_{\mathrm{opt}}}J(x;z),
\end{equation}
where $J(x;z)$ is instantiated by Speedbump-P or Speedbump-D, optionally combined with the target-behaviour regulariser (Section~\ref{sec:target_regularisation}). The optimisation runs for a fixed budget of $n_{\mathrm{iter}}$ iterations starting from a fixed initial suffix (Figure~\ref{fig:method_diagram}).

At each iteration, we refresh speculative rollouts under the incumbent suffix and freeze the resulting contexts and proposals while scoring candidates, so gradients do not pass through the discrete rollout. Rollouts are generated greedily regardless of the victim's decoding strategy, yielding deterministic candidate comparisons.

We compute gradients of $J$ with respect to the suffix-token representations, retain the $k$ tokens with the largest gradient entries at each position, and sample $n_{\mathrm{cand}}$ single-token substitutions. Candidates whose decoded text does not re-tokenise to the same token sequence are discarded. The remaining candidates and incumbent are rescored on the frozen rollout, and the best candidate is adopted only if it strictly improves the objective.

Because $J$ is a surrogate for realised speculative performance, we evaluate validation MAL (Equation~\ref{eq:MAL}) on $\mathcal{D}_{\mathrm{val}}$ before optimisation and every $n_{\mathrm{val\text{-}iter}}$ iterations, returning the checkpoint with the lowest validation MAL. No early stopping is applied.

\subsection{Target-Distribution Regularisation}
\label{sec:target_regularisation}

Our default attacks optimise the Speedbump objective alone. To study whether degradation can be retained while limiting changes to the target's behaviour, we also consider a variant that adds a regularisation term penalising shifts in the target distribution induced by the suffix. This variant is used only in Section~\ref{sec:stealthiness}; all other results use the unregularised objective.

For a prompt $x$, let $u_c$ denote the committed output prefix at the start of cycle $c$ in the current adversarial rollout. We evaluate the target model at the same output prefix with and without the suffix:
\begin{equation}
    p^{\mathrm{clean}}_c(\cdot)
    =
    P_T(\cdot \mid x, u_c),
    \qquad
    p^{z}_c(\cdot)
    =
    P_T(\cdot \mid x^z, u_c).
\end{equation}
Using the same committed prefix in both cases provides a matched comparison of the target distributions without introducing additional differences from the conditioning output history.

We define the regularisation term as the clean-to-suffixed KL divergence, averaged over the first $K_x=\min(20,C_x)$ cycle-base contexts:
\begin{equation}
    \mathcal{L}_{\mathrm{sem}}(x)
    =
    \frac{1}{K_x}
    \sum_{c=1}^{K_x}
    D_{\mathrm{KL}}
    \left(
        p^{\mathrm{clean}}_c
        \,\middle\|\,
        p^{z}_c
    \right).
    \label{eq:semantic_regularisation}
\end{equation}

This penalty can be combined with either Speedbump-P or Speedbump-D. For an attack objective $J_{\mathrm{Speedbump}}$, the regularised objective is
\begin{equation}
    J_{\beta}(x)
    =
    J_{\mathrm{Speedbump}}(x)
    -
    \beta
    \mathcal{L}_{\mathrm{sem}}(x),
    \label{eq:regularised_speedbump}
\end{equation}
where $\beta \geq 0$ controls the trade-off between speculative rejection and preservation of the clean target distribution. 

\section{Experimental Setup}

\subsection{Tasks and Datasets}
\label{sec:tasks_and_datasets}

We select a range of representative datasets covering mathematical reasoning, code generation, and text summarisation tasks. All datasets are obtained from HuggingFace (\texttt{openai/gsm8k}, \texttt{abisee/cnn\_dailymail}, \texttt{openai/openai\_humaneval}). For each dataset, we use fixed disjoint random sets of prompts for optimisation, validation, and evaluation.

\textit{GSM8K} \cite{gsm8k_cobbe2021trainingverifierssolvemath} is a dataset of 8.5K grade school math word problems. The task is to solve the problem, with accuracy judged by exact match between the final numerical answer and the reference answer. Optimisation and validation problems are drawn from the train split, and we evaluate on all 1,319 problems of the test split.

\textit{HumanEval} \cite{humaneval_chen2021evaluating} is a dataset of 164 simple Python programming problems. The task is to complete a function stub from its signature and docstring, with accuracy judged by pass@1 - the fraction of problems for which the model's completion passes all of the problem's unit tests. As HumanEval has a single split, we hold out 48 problems from which we draw the optimisation and validation prompts, and evaluate on the remaining 116.

\textit{CNN/DailyMail} \cite{cnndm, cnndm_see-etal-2017-get} (CNN/DM) is a dataset of 300k news articles written by journalists at CNN and the Daily Mail. We use version 3.0.0. The task is to summarise the article, with the quality of the summary judged by the ROUGE-L \cite{lin-2004-rouge} F-measure against the reference summary. After excluding articles that do not fit in the context window, optimisation and validation articles are drawn from the validation split, and we evaluate on 200 fixed random test articles.

\subsection{Speculative Decoding Methods}

We study 7 representative speculative decoding methods in our experiments. We selected from the most popular and effective speculative decoding methods, while covering the major drafter families: independent drafter (SpS \cite{leviathan2023fast,chen2023accelerating}), draft heads (Medusa \cite{medusa}, Hydra \cite{hydra}), feature-level drafters (EAGLE \cite{eagle}, EAGLE-2 \cite{eagle2}, EAGLE-3 \cite{eagle3}), and retrieval (PLD \cite{pld}).

Speculative Sampling (SpS) uses a separate lightweight drafter to autoregressively construct a single proposal chain. Medusa constructs tree-structured candidates from parallel draft heads, while Hydra additionally conditions later head predictions on earlier proposed tokens. Within the EAGLE family, EAGLE uses autoregressive feature-level drafting, EAGLE-2 introduces dynamic draft-tree construction, and EAGLE-3 performs direct token prediction from fused multi-layer target features. PLD serves as a model-free retrieval baseline, copying matching continuations from the input.

When combined with sampling, SpS and the EAGLE family use exact speculative sampling, which preserves the target distribution. Medusa and Hydra use typical acceptance, which accepts any proposal to which the target assigns sufficient probability and does not preserve the target distribution.

\subsection{Attack Evaluation}
\label{sec:attack_evaluation}

\subsubsection{Attack Effectiveness} We report two metrics for understanding the effectiveness of the attack: Mean Accepted Length (MAL) and speed-up.
MAL, as defined in Equation \ref{eq:MAL}, is the mean accepted length per cycle. As each target model forward-pass generates at least one verified token, MAL is bounded below by 1. As MAL depends only on the decoder's acceptance behaviour, and not the hardware, we rely on it as our primary metric.
Speedup is the ratio of speculative to autoregressive decoding throughput, where throughput is measured in generated tokens per second.
Clean speed-up is the tokens per second ratio of speculative decoding versus autoregressive decoding on the clean prompt, whereas attacked speed-up is the tokens per second ratio of speculative decoding versus autoregressive decoding on the suffixed prompt.
Speedup captures the wall-clock cost the defender pays, including the per-cycle cost of drafting and verification. A speed-up of smaller than 1x would mean that the speculative decoding is in fact \textit{slower} than the autoregressive decoding.
To control for hardware and load differences between machines, we time speculative and autoregressive decoding in succession on the same GPU.
MAL is measured per cycle and speed-up per generated token, so neither is lowered simply by an attack lengthening the output.

\subsubsection{Attack Stealthiness} We report two metrics which we use as proxies for the stealthiness of our attacks. Large changes in output distribution may be detectable automatically or manually. Firstly, we measure the output quality metric of each task to measure output quality, as mentioned in Section \ref{sec:tasks_and_datasets}. Secondly, in Section~\ref{sec:stealthiness} we additionally measure the perplexity of the output, conditioned on the prompt \textit{without} the suffix. This measures how natural the attacked output is as a response to the original request. Perplexity captures naturalness rather than correctness: a response can be fluent but wrong, or correct but awkwardly phrased.

\subsection{Implementation}
\label{sec:implementation}

\subsubsection{Models} We select Vicuna-13B v1.3 and Vicuna-7B v1.3 as target models - alongside Vicuna-33B, these are the only models which have pretrained heads for EAGLE, Medusa, and Hydra.
For Medusa, Hydra, and the EAGLE family, we use the draft weights released with the corresponding papers for these targets (note that no EAGLE-3 weights exist for Vicuna-7B). For SpS, we use the Vicuna-68M introduced by Multi-Candidate Speculative Decoding \cite{mcsd_10.1007/978-981-95-3343-5_26} as the drafter.

\subsubsection{Speculative Decoding} We based our implementations on SpecBench \cite{specbench-xia-etal-2024-unlocking}, a unified benchmark of different speculative decoding approaches. Every method uses the default SpecBench configuration, with two exceptions. Firstly, SpS drafts $\gamma=5$ tokens per cycle, rather than adapting $\gamma$ during generation according to the acceptance rate. Adapting $\gamma$ during generation would cause the draft budget itself to respond to the attack and confound the measured effect. Secondly, EAGLE-3 uses the same dynamic-tree budget as EAGLE-2 to facilitate easier comparison between the two methods. All experiments use batch size 1 and \texttt{float16} weights.

\subsubsection{Prompts and Generation} The prompting format and evaluation code are taken from EleutherAI's Language Model Evaluation Harness \cite{lmeval_lintang_sutawika_2023_10256836} for GSM8K (without the fewshot prompts) and CNN/DM, and from EvalPlus \cite{evalplus} for HumanEval. The exact prompt templates including the positions of our injected adversarial suffixes can be seen in Appendix \ref{appendix:prompts}. We limit generation to 1024 output tokens for the HumanEval and GSM8K datasets, and 512 tokens for the CNN/DM dataset, within the target models' 2048-token context window.

\subsubsection{Suffix Search} We use a suffix of length $L=20$ tokens, where each token is initialised to ``!". We optimise against $n_{opt} = 12$ optimisation prompts, and every $n_{\mathrm{val\text{-}iter}} = 10$ iterations we validate the suffix against $n_{val} = 24$ validation prompts. The optimisation is run for $n_{iter} = 400$ iterations. At each iteration, we retain the top $k = 256$ replacements per token, and sample $n_{cand} = 128$ single-token substitutions as candidate replacements. Before sampling the candidate replacements, we first filter out all candidates that tokenise differently from the original tokenisation when they are tokenised and re-tokenised on the optimisation prompts. Additionally, we restrict the first token to begin with a space to prevent the suffix fusing with the content it is placed after, and we prevent tokens with single quotation (') or double quotation marks (") to not break the parsing of the HumanEval docstring.

\subsubsection{Hardware} All experiments are performed on Red Hat Enterprise Linux 9.8 with an NVIDIA A100-SXM4-80GB GPU, AMD EPYC 7763 core, and 32GB DDR4-3200 RAM. All experiments are repeated for three seeds, with the reported results being the mean over those three seeds. For the sampling experiments, we vary the seed for the attack while keeping the seed for the sampling constant, to avoid confounding the variance of sampling with the variance of the attack.

\section{Results and Analysis}

We evaluate Speedbump along four axes: its effectiveness against greedy (Section~\ref{sec:attacking_greedy_decoding}) and sampled (Section~\ref{sec:attacking_sampling}) speculative decoding, the trade-off between degradation and stealthiness (Section~\ref{sec:stealthiness}), the contribution of each component of the objective (Section~\ref{sec:ablations}), and whether the suffixes remain effective when the victim's configuration differs from the one they were optimised against (Section~\ref{sec:attack_generalisability}).

\subsection{Attacking Greedy Decoding}
\label{sec:attacking_greedy_decoding}

Table \ref{tab:main_results_greedy_13b} reports clean and attacked speed-up, MAL, and task quality for Vicuna-13B under greedy decoding. Results for Vicuna-7B follow the same pattern (Appendix \ref{appendix:vicuna-7b}). PLD exposes no draft distribution, so we attack it only with Speedbump-P.

No decoder is robust: averaged over tasks, Speedbump-P reduces MAL by 19–31\% and Speedbump-D by 9–19\% for every decoder. The reduction isn't caused simply by appending tokens: 20-token random suffixes drawn from the same token pool change MAL by at most 0.04 on GSM8K and CNN/DM (Appendix \ref{appendix:per_seed_greedy}). The effect is largest on CNN/DM, where Speedbump-P reduces MAL by 42.6\% on average, and cuts EAGLE-3's MAL from 5.68 to 1.77. On GSM8K both attacks reduce MAL by roughly 19\%. HumanEval is the most resistant task, with an average MAL reduction of about 10\%. Random suffixes lower MAL on HumanEval by up to 0.31, and attacks on several tree-based methods (EAGLE-3/P, Medusa/P, Hydra/D) are no stronger than a random suffix. One possible explanation is that code is more predictable, leaving less scope to induce draft--target divergence.

Averaged over decoders, Speedbump-P and Speedbump-D achieve similar MAL reductions on GSM8K (18.5\% vs 19.0\%) and HumanEval (10.1\% vs 10.3\%). On CNN/DM, Speedbump-P is much stronger (42.6\% vs 16.8\%). Per decoder, the best attack varies. Speedbump-P is thus competitive with the distribution-aware attack, and clearly stronger on CNN/DM, while requiring no access to the drafter's distribution,

Speedup falls roughly in proportion to MAL, accompanied by a comparable loss in the task metric. For example, Speedbump-P's larger MAL reduction on CNN/DM (42.6\% vs 16.8\% for Speedbump-D) is matched by a larger reduction in ROUGE-L (47.7\% vs 18.5\%). Figure \ref{fig:effectiveness_stealthiness_tradeoff} examines whether the output quality can be maintained while retaining the degradation.

Per-seed results are given in Appendix \ref{appendix:per_seed_greedy}. Although no mean speed-up falls below 1$\times$, individual suffixes do, and several configurations show substantial variation across optimisation seeds. The reported means may therefore understate what an attacker can achieve by running multiple optimisations and selecting the suffix with the lowest validation MAL.

\begin{table*}[t]
\centering
\footnotesize
\caption{Attack results on Vicuna-13B across speculative decoding methods and datasets for greedy decoding. Each entry is formatted as clean/attacked. The final row reports the average absolute reduction marked by ↓, with the average relative reduction shown in parentheses.}
\label{tab:main_results_greedy_13b}
\setlength{\tabcolsep}{3pt}
\resizebox{\textwidth}{!}{%
\begin{tabular}{llccccccccc}
\toprule
& & \multicolumn{3}{c}{GSM8K} & \multicolumn{3}{c}{HumanEval} & \multicolumn{3}{c}{CNN/DM} \\
\cmidrule(lr){3-5} \cmidrule(lr){6-8} \cmidrule(lr){9-11}
Method & Attack & Speed-up$\downarrow$ & MAL$\downarrow$ & Acc. & Speed-up$\downarrow$ & MAL$\downarrow$ & Pass@1 & Speed-up$\downarrow$ & MAL$\downarrow$ & Rouge-L \\
\midrule
\multirow{2}{*}{SPS} & Speedbump-P & 1.54$\times$/1.20$\times$ & 2.38/1.88 & 26.5/11.4 & 2.10$\times$/1.62$\times$ & 2.75/2.28 & 14.7/11.2 & 1.57$\times$/1.13$\times$ & 2.44/1.65 & 24.3/14.0 \\
 & Speedbump-D & 1.54$\times$/1.34$\times$ & 2.38/2.04 & 26.5/22.1 & 2.10$\times$/1.50$\times$ & 2.75/1.88 & 14.7/11.8 & 1.57$\times$/1.56$\times$ & 2.44/2.29 & 24.3/23.9 \\
\addlinespace
\multirow{2}{*}{Medusa} & Speedbump-P & 2.02$\times$/1.41$\times$ & 2.65/1.78 & 26.5/15.0 & 2.35$\times$/1.98$\times$ & 2.70/2.61 & 14.7/16.1 & 1.52$\times$/1.05$\times$ & 2.11/1.34 & 24.2/8.6 \\
 & Speedbump-D & 2.02$\times$/1.43$\times$ & 2.65/1.90 & 26.5/13.0 & 2.35$\times$/2.05$\times$ & 2.70/2.53 & 14.7/14.9 & 1.52$\times$/1.40$\times$ & 2.11/1.80 & 24.2/19.9 \\
\addlinespace
\multirow{2}{*}{Hydra} & Speedbump-P & 2.50$\times$/2.19$\times$ & 3.74/3.31 & 26.5/22.0 & 2.58$\times$/2.38$\times$ & 3.86/3.52 & 14.7/14.9 & 1.92$\times$/1.23$\times$ & 2.88/1.69 & 24.2/12.1 \\
 & Speedbump-D & 2.50$\times$/1.66$\times$ & 3.74/2.39 & 26.5/13.1 & 2.58$\times$/2.47$\times$ & 3.86/3.64 & 14.7/15.8 & 1.92$\times$/1.63$\times$ & 2.88/2.45 & 24.2/18.4 \\
\addlinespace
\multirow{2}{*}{EAGLE} & Speedbump-P & 2.62$\times$/2.06$\times$ & 4.03/3.28 & 26.8/17.9 & 2.76$\times$/2.57$\times$ & 4.27/4.04 & 14.7/15.2 & 2.13$\times$/1.40$\times$ & 3.39/1.92 & 24.2/10.9 \\
 & Speedbump-D & 2.62$\times$/2.18$\times$ & 4.03/3.53 & 26.8/16.6 & 2.76$\times$/2.67$\times$ & 4.27/4.08 & 14.7/16.4 & 2.13$\times$/1.95$\times$ & 3.39/3.06 & 24.2/20.9 \\
\addlinespace
\multirow{2}{*}{EAGLE-2} & Speedbump-P & 3.06$\times$/2.47$\times$ & 4.97/4.09 & 26.5/12.9 & 3.11$\times$/2.95$\times$ & 5.28/4.80 & 14.7/15.5 & 2.42$\times$/1.76$\times$ & 4.08/2.63 & 24.2/15.6 \\
 & Speedbump-D & 3.06$\times$/2.69$\times$ & 4.97/4.28 & 26.5/16.3 & 3.11$\times$/3.01$\times$ & 5.28/4.96 & 14.7/6.3 & 2.42$\times$/1.98$\times$ & 4.08/3.06 & 24.2/17.8 \\
\addlinespace
\multirow{2}{*}{EAGLE-3} & Speedbump-P & 3.44$\times$/2.74$\times$ & 5.93/4.60 & 26.5/17.6 & 3.60$\times$/3.46$\times$ & 6.36/6.02 & 14.7/12.6 & 3.40$\times$/1.42$\times$ & 5.68/1.77 & 24.2/9.7 \\
 & Speedbump-D & 3.44$\times$/2.93$\times$ & 5.93/5.39 & 26.5/17.2 & 3.60$\times$/3.35$\times$ & 6.36/5.86 & 14.7/9.8 & 3.40$\times$/2.26$\times$ & 5.68/3.96 & 24.2/17.5 \\
\addlinespace
PLD & Speedbump-P & 1.53$\times$/1.53$\times$ & 1.75/1.65 & 26.8/16.6 & 1.93$\times$/1.48$\times$ & 2.09/1.64 & 14.7/8.9 & 1.97$\times$/1.12$\times$ & 2.12/1.27 & 24.2/17.7 \\
\midrule
\multirow{2}{*}{Avg. drop} & Speedbump-P & {$\downarrow$0.44$\times$}\,{\scriptsize(18.0\%)} & {$\downarrow$0.69}\,{\scriptsize(18.5\%)} & {$\downarrow$10.4}\,{\scriptsize(39.1\%)} & {$\downarrow$0.28$\times$}\,{\scriptsize(12.2\%)} & {$\downarrow$0.34}\,{\scriptsize(10.1\%)} & {$\downarrow$1.2}\,{\scriptsize(8.3\%)} & {$\downarrow$0.83$\times$}\,{\scriptsize(36.8\%)} & {$\downarrow$1.49}\,{\scriptsize(42.6\%)} & {$\downarrow$11.6}\,{\scriptsize(47.7\%)} \\
 & Speedbump-D & {$\downarrow$0.49$\times$}\,{\scriptsize(19.9\%)} & {$\downarrow$0.69}\,{\scriptsize(19.0\%)} & {$\downarrow$10.2}\,{\scriptsize(38.3\%)} & {$\downarrow$0.24$\times$}\,{\scriptsize(9.8\%)} & {$\downarrow$0.38}\,{\scriptsize(10.3\%)} & {$\downarrow$2.2}\,{\scriptsize(15.0\%)} & {$\downarrow$0.36$\times$}\,{\scriptsize(14.0\%)} & {$\downarrow$0.66}\,{\scriptsize(16.8\%)} & {$\downarrow$4.5}\,{\scriptsize(18.5\%)} \\
\bottomrule
\end{tabular}}
\end{table*}

\subsection{Attacking Sampling}
\label{sec:attacking_sampling}

Speculative decoding is also used with sampling, under a modified acceptance rule (Section~\ref{sec:preliminaries}). Suffix search uses greedy rollouts regardless of the victim's decoding strategy (Section~\ref{sec:suffix_search}), so we evaluate the same suffixes at temperature 1 (Table \ref{tab:main_results_sampling_13b}). SpS and the EAGLE family use exact speculative sampling, and Medusa and Hydra use typical acceptance; PLD supports only greedy decoding and is excluded.

The attacks remain effective under sampling, with MAL reductions similar to those under greedy decoding, and again CNN/DM the most vulnerable and HumanEval the most resistant. Sampling lowers clean speed-up, and some attacks, particularly on CNN/DM, push speed-up below 1$\times$, making speculative decoding slower than autoregressive decoding (see Appendix \ref{appendix:per_seed_sampling} for per-seed results).

Speedbump-D's acceptance proxy is exactly the acceptance probability of speculative sampling, while Speedbump-P's acceptance proxy is a relaxation of greedy acceptance (Section~\ref{sec:acceptance_proxies}). Accordingly, for the four exact-sampling decoders, Speedbump-D's average MAL reduction rises from 14.2\% under greedy decoding to 18.7\%, while Speedbump-P's barely changes (24.7\% to 25.9\%). While Speedbump-P's acceptance proxy isn't designed for sampling, the proposed tokens are where the drafter concentrates its mass, so lowering their probabilities also reduces the draft-target overlap that governs sampled acceptance.

Under typical acceptance (Medusa and Hydra), both attacks weaken: Speedbump-P falls from 22.4\% to 18.2\% and Speedbump-D from 17.7\% to 12.6\%. Typical acceptance is more lenient than greedy verification: greedy verification rejects a proposal whenever it differs from the target's most likely token, while typical acceptance rejects it only if its target probability falls below a threshold. We hypothesise that suffixes which merely change the target's top token cause rejections under greedy verification but not under typical acceptance, and neither proxy targets the threshold directly.

\begin{table*}[t]
\centering
\footnotesize
\caption{Attack results on Vicuna-13B across speculative decoding methods and datasets for temperature-1 sampling. Each entry is formatted as clean/attacked. The final row reports the average absolute reduction marked by ↓, with the average relative reduction shown in parentheses. Methods marked $^\dagger$ perform typical acceptance.}
\label{tab:main_results_sampling_13b}
\setlength{\tabcolsep}{3pt}
\resizebox{\textwidth}{!}{%
\begin{tabular}{llccccccccc}
\toprule
& & \multicolumn{3}{c}{GSM8K} & \multicolumn{3}{c}{HumanEval} & \multicolumn{3}{c}{CNN/DM} \\
\cmidrule(lr){3-5} \cmidrule(lr){6-8} \cmidrule(lr){9-11}
Method & Attack & Speed-up$\downarrow$ & MAL$\downarrow$ & Acc. & Speed-up$\downarrow$ & MAL$\downarrow$ & Pass@1 & Speed-up$\downarrow$ & MAL$\downarrow$ & RL \\
\midrule
\multirow{2}{*}{SPS} & Speedbump-P & 1.28$\times$/0.98$\times$ & 1.99/1.56 & 20.4/5.5 & 1.38$\times$/1.26$\times$ & 2.11/1.91 & 10.3/7.2 & 1.32$\times$/0.90$\times$ & 2.12/1.45 & 23.3/11.7 \\
 & Speedbump-D & 1.28$\times$/1.04$\times$ & 1.99/1.62 & 20.4/12.6 & 1.38$\times$/1.13$\times$ & 2.11/1.69 & 10.3/11.8 & 1.32$\times$/1.18$\times$ & 2.12/1.83 & 23.3/22.8 \\
\addlinespace
\multirow{2}{*}{Medusa$^\dagger$} & Speedbump-P & 2.08$\times$/1.77$\times$ & 2.88/2.16 & 25.0/16.4 & 2.30$\times$/2.27$\times$ & 2.97/2.92 & 19.8/13.5 & 1.69$\times$/1.22$\times$ & 2.29/1.50 & 23.6/9.3 \\
 & Speedbump-D & 2.08$\times$/1.70$\times$ & 2.88/2.18 & 25.0/11.5 & 2.30$\times$/2.26$\times$ & 2.97/2.90 & 19.8/16.1 & 1.69$\times$/1.50$\times$ & 2.29/1.96 & 23.6/20.5 \\
\addlinespace
\multirow{2}{*}{Hydra$^\dagger$} & Speedbump-P & 2.74$\times$/2.46$\times$ & 4.02/3.65 & 24.0/21.7 & 2.76$\times$/2.78$\times$ & 4.12/3.89 & 12.9/14.9 & 2.16$\times$/1.58$\times$ & 3.20/2.14 & 22.8/12.5 \\
 & Speedbump-D & 2.74$\times$/2.18$\times$ & 4.02/3.34 & 24.0/9.9 & 2.76$\times$/2.81$\times$ & 4.12/4.02 & 12.9/15.5 & 2.16$\times$/1.79$\times$ & 3.20/2.72 & 22.8/17.9 \\
\addlinespace
\multirow{2}{*}{EAGLE} & Speedbump-P & 2.28$\times$/1.72$\times$ & 3.71/2.91 & 20.2/10.7 & 2.43$\times$/2.42$\times$ & 3.89/3.79 & 10.6/11.2 & 1.89$\times$/1.08$\times$ & 3.15/1.66 & 23.4/8.8 \\
 & Speedbump-D & 2.28$\times$/1.64$\times$ & 3.71/2.80 & 20.2/6.2 & 2.43$\times$/2.54$\times$ & 3.89/3.78 & 10.6/10.3 & 1.89$\times$/1.69$\times$ & 3.15/2.68 & 23.4/17.2 \\
\addlinespace
\multirow{2}{*}{EAGLE-2} & Speedbump-P & 2.74$\times$/2.02$\times$ & 4.54/3.50 & 20.4/7.7 & 2.90$\times$/2.95$\times$ & 4.83/4.49 & 12.1/7.8 & 2.15$\times$/1.36$\times$ & 3.84/2.26 & 23.3/12.2 \\
 & Speedbump-D & 2.74$\times$/2.03$\times$ & 4.54/3.48 & 20.4/10.3 & 2.90$\times$/2.71$\times$ & 4.83/4.38 & 12.1/4.9 & 2.15$\times$/1.42$\times$ & 3.84/2.44 & 23.3/11.8 \\
\addlinespace
\multirow{2}{*}{EAGLE-3} & Speedbump-P & 3.05$\times$/2.18$\times$ & 5.37/3.83 & 19.9/7.3 & 3.34$\times$/3.30$\times$ & 5.78/5.49 & 12.4/12.4 & 2.78$\times$/0.96$\times$ & 5.26/1.49 & 23.2/6.8 \\
 & Speedbump-D & 3.05$\times$/2.54$\times$ & 5.37/4.67 & 19.9/7.8 & 3.34$\times$/3.20$\times$ & 5.78/5.18 & 12.4/6.3 & 2.78$\times$/1.79$\times$ & 5.26/3.30 & 23.2/15.3 \\
\midrule
\multirow{2}{*}{Avg. drop} & Speedbump-P & {$\downarrow$0.51$\times$}\,{\scriptsize(21.3\%)} & {$\downarrow$0.82}\,{\scriptsize(21.5\%)} & {$\downarrow$10.1}\,{\scriptsize(48.3\%)} & {$\downarrow$0.02$\times$}\,{\scriptsize(1.5\%)} & {$\downarrow$0.20}\,{\scriptsize(5.2\%)} & {$\downarrow$1.8}\,{\scriptsize(12.7\%)} & {$\downarrow$0.81$\times$}\,{\scriptsize(38.6\%)} & {$\downarrow$1.56}\,{\scriptsize(43.2\%)} & {$\downarrow$13.0}\,{\scriptsize(56.0\%)} \\
 & Speedbump-D & {$\downarrow$0.51$\times$}\,{\scriptsize(21.4\%)} & {$\downarrow$0.74}\,{\scriptsize(20.1\%)} & {$\downarrow$11.9}\,{\scriptsize(55.1\%)} & {$\downarrow$0.08$\times$}\,{\scriptsize(4.0\%)} & {$\downarrow$0.29}\,{\scriptsize(7.9\%)} & {$\downarrow$2.2}\,{\scriptsize(15.9\%)} & {$\downarrow$0.44$\times$}\,{\scriptsize(19.9\%)} & {$\downarrow$0.82}\,{\scriptsize(22.0\%)} & {$\downarrow$5.7}\,{\scriptsize(24.4\%)} \\
\bottomrule
\end{tabular}}
\end{table*}

\subsection{Effectiveness-Stealthiness Tradeoff}
\label{sec:stealthiness}

The attacker's primary aim is to raise the victim's inference cost, so a loss in output quality need not concern them, and may even be welcome, as it degrades the victim's service further. A visible change in the output, however, makes the attack easier to detect. We therefore examine whether the degradation can be achieved while leaving the output intact.

\begin{figure*}
    \centering
    \includegraphics[width=0.85\linewidth]{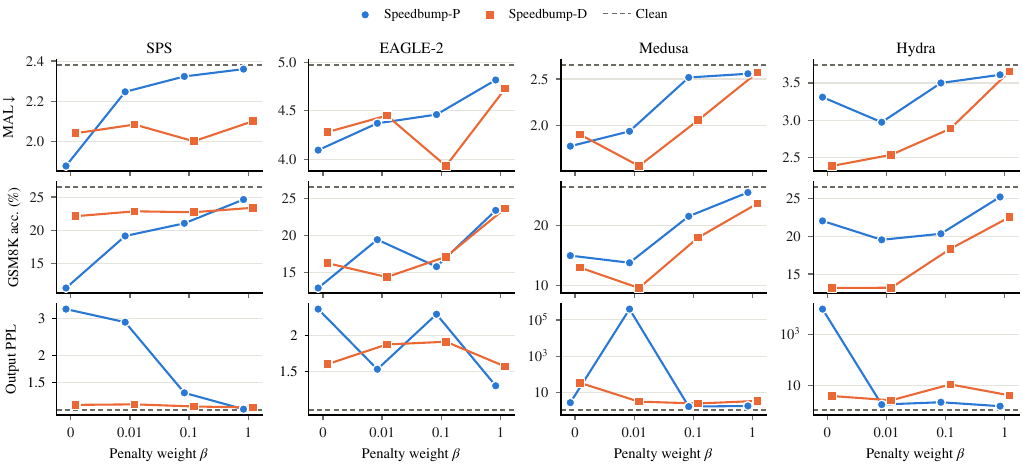}
    \caption{MAL, accuracy, and output perplexity (PPL) on GSM8K for different $\beta$ weightings of the target-distribution regulariser.}
    \label{fig:effectiveness_stealthiness_tradeoff}
\end{figure*}

If the degradation comes mainly from changing what the target generates, penalising that change should remove it. We test this with the target-distribution regulariser (Section~\ref{sec:target_regularisation}) on GSM8K for SpS, EAGLE-2, Medusa and Hydra, with $\beta \in \{0.01, 0.1, 1\}$ (Figure~\ref{fig:effectiveness_stealthiness_tradeoff}). As $\beta$ increases, the regulariser generally succeeds in improving attack stealthiness: accuracy recovers to near clean levels, and the output perplexity decreases. However, this comes at a cost to the attack effectiveness: MAL generally increases to near clean levels. Excluding SpS/D, the configurations recover 70--93\% of their accuracy loss between $\beta = 0$ and $\beta = 1$, but lose 66--95\% of their MAL reduction. The effect of $\beta$ is not monotonic: for Medusa/D and Hydra/P, the weakest penalty ($\beta = 0.01$) produces a larger degradation than the unregularised attack. The regulariser therefore trades effectiveness for stealthiness rather than finding a better trade-off, indicating that most of the degradation comes from changing what the target generates. This conclusion is limited to one penalty on one task, with checkpoints selected by validation MAL alone.

SpS/D is the exception, reducing MAL on GSM8K by 12--16\% at every $\beta$. Unlike the other drafters, which read the target's hidden states, SpS's drafter is a separate model that reads the suffix itself, so Speedbump-D can reduce the draft--target overlap by changing the drafter's predictions rather than the target's. This suggests that attacks acting on the drafter could slow decoding with less visible changes to the output.

\subsection{Ablations}
\label{sec:ablations}

We perform ablations to demonstrate the importance of the components of our objective.

\subsubsection{Objective}
\label{sec:objective_ablations}

We compare Speedbump with two objectives on the scored token positions. Acc removes the prefix reachability weighting, taking a flat mean of $\alpha_{c,i}$ instead of summing the prefix products of Equation \ref{eq:speedbump_objective}. Log further replaces $\alpha_{c,i}$ with a log-space score rather than a raw score. Each objective also has a -decay variant, which weights each scored context linearly by its token position, prioritising earlier tokens. P/Log corresponds to Mistletoe's \cite{sun2026mistletoestealthyaccelerationcollapseattacks} objective, and D/Log-decay to ADSD's \cite{ADSDwang2026adversarialpromptsacceptancecollapse} Soft-Collapse objective. We implement only their objectives while maintaining our search and threat model. Table \ref{tab:objective_ablation} reports the MAL for the different objectives.

We find that Speedbump outperforms both Mistletoe and ADSD's objectives on every decoder. A log-space score rewards reductions even on tokens that are already rejected, whereas $\alpha$ saturates once acceptance is unlikely. Weighting early output positions with a decay makes Speedbump substantially stronger, causing Speedbump-decay to be the strongest objective for SpS and Hydra. 

\begin{table}[t]
\centering
\footnotesize
\caption{MAL (lower is better) for different attack objectives of Speedbump-P (P) and Speedbump-D (D) on GSM8K (Vicuna-13B).
$^\dagger$ denotes the Mistletoe attack objective, $^\ddagger$ denotes the ADSD attack objective.}
\label{tab:objective_ablation}
\setlength{\tabcolsep}{3pt}
\begin{tabular}{lcccccc|cc}
\toprule
& \multicolumn{2}{c}{SPS} & \multicolumn{2}{c}{EAGLE-2} & \multicolumn{2}{c}{Hydra} & \multicolumn{2}{|c}{Avg.} \\
\cmidrule(lr){2-3} \cmidrule(lr){4-5} \cmidrule(lr){6-7} \cmidrule(lr){8-9}
Shape & P & D & P & D & P & D & P & D \\
\midrule
Clean & 2.38 & 2.38 & 4.97 & 4.97 & 3.74 & 3.74 & 3.70 & 3.70 \\
\midrule
Log & 2.36$^\dagger$ & 2.07 & 4.80$^\dagger$ & 4.52 & 3.53$^\dagger$ & 3.51 & 3.56$^\dagger$ & 3.37 \\
Log-decay & 2.35 & 2.20$^\ddagger$ & 4.73 & 4.61$^\ddagger$ & 3.40 & 3.46$^\ddagger$ & 3.49 & 3.42$^\ddagger$ \\
Acc & 2.04 & 2.10 & 4.49 & 3.51 & 2.37 & 2.31 & 2.97 & \textbf{2.64} \\
Acc-decay & 1.84 & 2.09 & \textbf{3.29} & \textbf{3.25} & 2.53 & 3.04 & 2.55 & 2.79 \\
Speedbump & 1.88 & \textbf{2.04} & 4.09 & 4.28 & 3.31 & 2.39 & 3.09 & 2.90 \\
Speedbump-decay & \textbf{1.60} & 2.06 & 3.33 & 3.94 & \textbf{1.97} & \textbf{2.23} & \textbf{2.30} & 2.74 \\
\bottomrule
\end{tabular}
\end{table}

\subsubsection{Score Depth}
\label{sec:score_depth_ablations}

Speedbump scores every depth reached in a cycle. We test whether the deeper depths contribute by capping the scored depths per cycle at 1 or 3. With a cap of 1, we only use the mean acceptance proxy at the first depth, with no prefix-reachability weighting at all. Table \ref{tab:depth_ablation} reports the MAL for different scoring depths. We find that scoring beyond the first depth helps. Scoring all depths, however, seems to give no further gain over depth 3. The deepest depths are the products of many proxies, and they are conditioned on the least reliable parts of the recorded rollout.

\begin{table}[t]
\centering
\footnotesize
\caption{MAL (lower is better) for different scoring depths of Speedbump-P (P) and Speedbump-D (D) on GSM8K (Vicuna-13B).}
\label{tab:depth_ablation}
\setlength{\tabcolsep}{4pt}
\begin{tabular}{lcccccc|cc}
\toprule
& \multicolumn{2}{c}{SPS} & \multicolumn{2}{c}{EAGLE-2} & \multicolumn{2}{c}{Hydra} & \multicolumn{2}{|c}{Avg.} \\
\cmidrule(lr){2-3} \cmidrule(lr){4-5} \cmidrule(lr){6-7} \cmidrule(lr){8-9}
Depth & P & D & P & D & P & D & P & D \\
\midrule
Clean & 2.38 & 2.38 & 4.97 & 4.97 & 3.74 & 3.74 & 3.70 & 3.70 \\
\midrule
1 & 2.11 & 2.00 & 4.69 & 4.79 & 3.19 & 3.30 & 3.33 & 3.36 \\
3 & \textbf{1.81} & \textbf{1.96} & 4.60 & \textbf{4.24} & \textbf{2.87} & \textbf{2.38} & \textbf{3.09} & \textbf{2.86} \\
All & 1.88 & 2.04 & \textbf{4.09} & 4.28 & 3.31 & 2.39 & \textbf{3.09} & 2.90 \\
\bottomrule
\end{tabular}
\end{table}

Overall, the ablations support what distinguishes Speedbump from prior objectives. Its bounded acceptance proxies outperform the Mistletoe- and ADSD-style objectives on every decoder, and scoring several depths per cycle outperforms scoring only the first. They also show that the objective leaves room for improvement. Weighting early output positions more increases the MAL reduction of our attack, while capping the score depth at 3 does as well as the full prefix product. Both results suggest that an attack objective is most effective when it focuses on the positions that shape the rest of the output, early in each cycle and early in the rollout. Our main results are therefore a conservative estimate of what this attack surface allows. With one task, three decoders and substantial seed variation, however, the ablations do not establish a general ranking of objectives.

\subsection{Attack Generalisability}
\label{sec:attack_generalisability}

Our attacks assume white-box access to the target model, as well as access to either the full draft distribution (Speedbump-D) or just the draft proposals (Speedbump-P). In practice, an attacker may not know or have access to the victim's target model or drafter. This section studies whether the attack remains effective when the victim configuration differs from the one it was optimised against.

\subsubsection{Cross-Drafter Transferability}
\label{sec:cross_drafter_transferability}

A victim may serve a known open-weight target with an undisclosed or custom-trained drafter. We study whether suffixes optimised against one drafter transfer to other drafters. We evaluate the suffixes optimised against SpS on Vicuna-13B on the six other decoders, with the same target model and task. Table \ref{tab:cross_draft} compares each transferred suffix (SpS) with the suffix optimised directly against the destination decoder (Nat.).

Speedbump-P transfers strongly across drafters. On GSM8K, the SpS/P suffixes reduce MAL on the other decoders by 1.20 on average, more than the native suffixes (0.73), and on CNN/DM they retain about 70\% of the native reduction. Transfer is weak only on HumanEval. The SpS/P suffixes transfer well because they act on the target's output rather than on the drafter, remaining drafter-agnostic. Speedbump-D does not transfer across drafters: on GSM8K and CNN/DM, the SpS/D suffixes change the MAL of every other decoder by at most 0.05.

\begin{table*}[t]
\centering
\footnotesize
\caption{Cross-draft transfer: MAL (lower is stronger) of suffixes optimised against SPS and transfered to other drafters, on Vicuna-13B. Clean denotes no suffix, Nat. denotes the suffix optimised directly against the drafter, and SPS denotes the suffix transfered from SPS.}
\label{tab:cross_draft}
\setlength{\tabcolsep}{4pt}
\resizebox{\ifdim\width>\textwidth\textwidth\else\width\fi}{!}{%
\begin{tabular}{llccccccccc}
\toprule
& & \multicolumn{3}{c}{GSM8K} & \multicolumn{3}{c}{HumanEval} & \multicolumn{3}{c}{CNN/DM} \\
\cmidrule(lr){3-5} \cmidrule(lr){6-8} \cmidrule(lr){9-11}
Decoder & Attack & Clean & Nat. & SPS & Clean & Nat. & SPS & Clean & Nat. & SPS \\
\midrule
\multirow{2}{*}{SPS (source)} & Speedbump-P & 2.38 & 1.88 & -- & 2.75 & 2.28 & -- & 2.44 & 1.65 & -- \\
 & Speedbump-D & 2.38 & 2.04 & -- & 2.75 & 1.88 & -- & 2.44 & 2.29 & -- \\
\midrule
\multirow{2}{*}{Medusa} & Speedbump-P & 2.65 & 1.78 & 1.98 & 2.70 & 2.61 & 2.70 & 2.11 & 1.34 & 1.68 \\
 & Speedbump-D & 2.65 & 1.90 & 2.65 & 2.70 & 2.53 & 2.38 & 2.11 & 1.80 & 2.11 \\
\addlinespace
\multirow{2}{*}{Hydra} & Speedbump-P & 3.74 & 3.31 & 2.61 & 3.86 & 3.52 & 3.76 & 2.88 & 1.69 & 2.13 \\
 & Speedbump-D & 3.74 & 2.39 & 3.72 & 3.86 & 3.64 & 3.36 & 2.88 & 2.45 & 2.88 \\
\addlinespace
\multirow{2}{*}{EAGLE} & Speedbump-P & 4.03 & 3.28 & 2.79 & 4.27 & 4.04 & 4.12 & 3.39 & 1.92 & 2.28 \\
 & Speedbump-D & 4.03 & 3.53 & 4.00 & 4.27 & 4.08 & 4.15 & 3.39 & 3.06 & 3.38 \\
\addlinespace
\multirow{2}{*}{EAGLE-2} & Speedbump-P & 4.97 & 4.09 & 3.30 & 5.28 & 4.80 & 5.10 & 4.08 & 2.63 & 2.66 \\
 & Speedbump-D & 4.97 & 4.28 & 4.94 & 5.28 & 4.96 & 5.08 & 4.08 & 3.06 & 4.08 \\
\addlinespace
\multirow{2}{*}{EAGLE-3} & Speedbump-P & 5.93 & 4.60 & 3.71 & 6.36 & 6.02 & 6.18 & 5.68 & 1.77 & 3.26 \\
 & Speedbump-D & 5.93 & 5.39 & 5.88 & 6.36 & 5.86 & 5.88 & 5.68 & 3.96 & 5.67 \\
\addlinespace
\multirow{2}{*}{PLD} & Speedbump-P & 1.75 & 1.65 & 1.51 & 2.09 & 1.64 & 1.75 & 2.12 & 1.27 & 1.43 \\
 & Speedbump-D & 1.75 & -- & 1.76 & 2.09 & -- & 2.26 & 2.12 & -- & 2.09 \\
\midrule
\multirow{2}{*}{Avg. drop} & Speedbump-P &  & $\downarrow$0.73 & $\downarrow$1.20 &  & $\downarrow$0.32 & $\downarrow$0.16 &  & $\downarrow$1.61 & $\downarrow$1.14 \\
 & Speedbump-D &  & $\downarrow$0.77 & $\downarrow$0.03 &  & $\downarrow$0.28 & $\downarrow$0.32 &  & $\downarrow$0.76 & $\downarrow$0.00 \\
\bottomrule
\end{tabular}}
\end{table*}

\subsubsection{Cross-Target-Model Transferability}
\label{sec:cross_target_transferability}

We study whether suffixes optimised against one target model transfer to another. We evaluate the suffixes optimised against Vicuna-13B on Vicuna-7B, and vice versa, keeping the decoding method and task fixed. As EAGLE-3 has no Vicuna-7B checkpoint, it is excluded here. The drafters of Medusa, Hydra and the EAGLE family are trained for a specific target, so changing the target model also changes the drafter. The exceptions are SpS, which uses the same drafter for both targets, and PLD, which has none. Table \ref{tab:cross_target} compares the MAL of each transferred suffix (Tr.) with the suffix optimised directly against the destination target (Nat.).

Speedbump-P transfers poorly across target models. Averaged over decoders, transferred Speedbump-P suffixes retain between 12\% and 43\% of the native MAL reduction. Most of Speedbump-P's degradation comes from changing what the target generates (Section~\ref{sec:stealthiness}), and this change largely does not carry over to the other target. A suffix that changes the output of one target model therefore does not reliably have the same effect on another, even within the same model family.

Speedbump-D transfers better than Speedbump-P. For SpS, the suffixes transfer almost completely on all three tasks, and in three of the six cases it outperforms the native one. SpS is the only neural decoding method whose drafter is identical for both targets, so a suffix that acts on the drafter (Section~\ref{sec:stealthiness}) keeps its effect regardless of which target the drafter serves. For the other decoders, Speedbump-D transfers inconsistently, and transfers poorly for the CNN/DM dataset. For these decoders, both the draft and target distributions change, and a suffix tuned to lower the overlap of one pair struggles to lower that of another.

The two attack methods therefore transfer in complementary ways. Speedbump-P's suffixes, that tend to change the target's output, transfer well between drafters while the target remains fixed, but not between targets. Speedbump-D's suffixes, that tend to change the drafter's output, transfer well between targets while the drafter remains fixed, but not between drafters.

\begin{table*}[t]
\centering
\footnotesize
\caption{Cross-target transfer: MAL (lower is stronger) of suffixes optimised against Vicuna-13B and transferred to Vicuna-7B, and vice versa. Clean denotes no suffix, Nat. denotes the suffix optimised directly against the target model, and Tr. denotes the suffix transfered from the other target model.}
\label{tab:cross_target}
\setlength{\tabcolsep}{2pt}
\resizebox{\ifdim\width>\textwidth\textwidth\else\width\fi}{!}{%
\begin{tabular}{llcccccccccccccccccc}
\toprule
& & \multicolumn{9}{c}{Vicuna-13B $\rightarrow$ 7B} & \multicolumn{9}{c}{Vicuna-7B $\rightarrow$ 13B} \\
\cmidrule(lr){3-11} \cmidrule(lr){12-20}
& & \multicolumn{3}{c}{GSM8K} & \multicolumn{3}{c}{HumanEval} & \multicolumn{3}{c}{CNN/DM} & \multicolumn{3}{c}{GSM8K} & \multicolumn{3}{c}{HumanEval} & \multicolumn{3}{c}{CNN/DM} \\
\cmidrule(lr){3-5} \cmidrule(lr){6-8} \cmidrule(lr){9-11} \cmidrule(lr){12-14} \cmidrule(lr){15-17} \cmidrule(lr){18-20}
Method & Attack & Clean & Nat. & Tr. & Clean & Nat. & Tr. & Clean & Nat. & Tr. & Clean & Nat. & Tr. & Clean & Nat. & Tr. & Clean & Nat. & Tr. \\
\midrule
\multirow{2}{*}{SPS} & Speedbump-P & 2.47 & 2.44 & 2.22 & 2.62 & 2.10 & 2.57 & 2.49 & 1.77 & 2.43 & 2.38 & 1.88 & 2.32 & 2.75 & 2.28 & 2.70 & 2.44 & 1.65 & 2.04 \\
 & Speedbump-D & 2.47 & 1.92 & 2.05 & 2.62 & 2.06 & 2.01 & 2.49 & 2.43 & 2.37 & 2.38 & 2.04 & 1.93 & 2.75 & 1.88 & 2.10 & 2.44 & 2.29 & 2.32 \\
\addlinespace
\multirow{2}{*}{Medusa} & Speedbump-P & 2.59 & 1.67 & 2.30 & 2.79 & 2.67 & 2.73 & 2.03 & 1.62 & 1.85 & 2.65 & 1.78 & 2.65 & 2.70 & 2.61 & 2.67 & 2.11 & 1.34 & 1.95 \\
 & Speedbump-D & 2.59 & 2.13 & 2.28 & 2.79 & 2.62 & 2.74 & 2.03 & 1.86 & 2.03 & 2.65 & 1.90 & 2.62 & 2.70 & 2.53 & 2.68 & 2.11 & 1.80 & 2.05 \\
\addlinespace
\multirow{2}{*}{Hydra} & Speedbump-P & 3.70 & 2.94 & 3.50 & 3.92 & 3.64 & 3.80 & 2.77 & 1.89 & 2.37 & 3.74 & 3.31 & 3.67 & 3.86 & 3.52 & 3.72 & 2.88 & 1.69 & 2.52 \\
 & Speedbump-D & 3.70 & 2.76 & 3.08 & 3.92 & 3.70 & 3.91 & 2.77 & 2.63 & 2.76 & 3.74 & 2.39 & 3.48 & 3.86 & 3.64 & 3.73 & 2.88 & 2.45 & 2.85 \\
\addlinespace
\multirow{2}{*}{EAGLE} & Speedbump-P & 4.04 & 3.69 & 3.93 & 4.22 & 4.04 & 4.22 & 3.29 & 2.36 & 2.81 & 4.03 & 3.28 & 4.00 & 4.27 & 4.04 & 4.13 & 3.39 & 1.92 & 2.97 \\
 & Speedbump-D & 4.04 & 2.61 & 3.93 & 4.22 & 4.19 & 4.24 & 3.29 & 3.24 & 3.21 & 4.03 & 3.53 & 3.28 & 4.27 & 4.08 & 4.17 & 3.39 & 3.06 & 3.36 \\
\addlinespace
\multirow{2}{*}{EAGLE-2} & Speedbump-P & 5.02 & 4.05 & 4.84 & 5.36 & 5.18 & 5.31 & 3.96 & 2.29 & 3.41 & 4.97 & 4.09 & 4.34 & 5.28 & 4.80 & 5.07 & 4.08 & 2.63 & 3.42 \\
 & Speedbump-D & 5.02 & 3.36 & 4.36 & 5.36 & 5.25 & 5.42 & 3.96 & 3.52 & 4.03 & 4.97 & 4.28 & 4.26 & 5.28 & 4.96 & 5.20 & 4.08 & 3.06 & 4.07 \\
\addlinespace
PLD & Speedbump-P & 1.85 & 1.81 & 1.61 & 1.82 & 1.52 & 1.90 & 2.24 & 1.36 & 1.59 & 1.75 & 1.65 & 1.75 & 2.09 & 1.64 & 1.98 & 2.12 & 1.27 & 1.92 \\
\midrule
\multirow{2}{*}{Avg. drop} & Speedbump-P &  & $\downarrow$0.51 & $\downarrow$0.21 &  & $\downarrow$0.26 & $\downarrow$0.03 &  & $\downarrow$0.91 & $\downarrow$0.39 &  & $\downarrow$0.59 & $\downarrow$0.13 &  & $\downarrow$0.34 & $\downarrow$0.11 &  & $\downarrow$1.09 & $\downarrow$0.37 \\
 & Speedbump-D &  & $\downarrow$1.01 & $\downarrow$0.42 &  & $\downarrow$0.22 & $\downarrow$0.12 &  & $\downarrow$0.17 & $\downarrow$0.03 &  & $\downarrow$0.73 & $\downarrow$0.44 &  & $\downarrow$0.35 & $\downarrow$0.20 &  & $\downarrow$0.45 & $\downarrow$0.05 \\
\bottomrule
\end{tabular}}
\end{table*}

\section{Discussion}

\subsection{Scope, Limitations, and Threats to Validity}
Generation prolonging attacks affect both autoregressive decoding and speculative decoding, and are limited only by a possible output-length cap. By contrast, SRAs do not target vanilla autoregressive decoding, and can at most remove the speculative speed-up while still incurring the per-cycle drafting and verification overhead, leading to generation slightly slower than vanilla decoding.

Our attacks require white-box access to the target model during optimisation, and Speedbump-P and Speedbump-D additionally require access to the drafter's proposals or distributions respectively. Suffixes transfer to some extent across drafters, so the attacker can use a surrogate drafter. However, transfer across target models is poor, even between Vicuna models sharing a tokeniser. In practice, the attack therefore requires the attacker to know, or correctly guess, the victim's target model. This is plausible for applications built on widely used open-weight models, but the attack is unlikely to be effective against closed or unknown models without improved transferability. We use only a single prompt template per task. Thus, it remains to be seen how robust suffixes are to prompt variations for the same task. Similarly, our suffixes are always optimised on greedy rollouts, whereas optimising on sampled rollouts could yield suffixes better suited to victims that sample.

We evaluate our attacks using SpecBench, a framework designed for research, as well as using a batch size of 1. Production serving stacks such as vLLM \cite{vllm} and SGLang \cite{sglang} use continuous batching, which shifts the workload from memory-bandwidth-bound to compute-bound. This reduces the speed-up of speculative decoding, which gives our attack less to degrade. Conversely, compute spent verifying rejected drafts is shared with co-batched requests, so an attack could also reduce throughput for other users, an effect our single-request setting does not capture.

The metrics we use to judge the quality of the response, and thus the stealthiness of the attacks, are coarse. ROUGE-L is a weak proxy for summary quality, and exact-match and pass@1 only capture task correctness, not whether outputs look different to a user. Perplexity captures fluency, but not correctness. In cases where the SRA has hijacked the output, manual inspection of the output could thus expose the presence of an SRA.

\subsection{Countermeasures and Mitigations}
One common defence against adversarial perturbations is to detect an attack by the perplexity, as the unnatural suffix may raise the perplexity of the prompt \cite{alon2023detectinglanguagemodelattacks}. While somewhat effective, such a defence may be bypassed with an adaptive attack which adds a fluency term to the objective \cite{10.1145/3128572.3140444, 3495724.3495862, zhu2024autodan}.

Further defences may attempt to detect an attack from the generated response, for example by detecting an increase in response perplexity, or a drop in response quality or mean accepted length. Detecting a drop in response quality or mean accepted length relies on having a reliable estimate of the expected task metric, as well as collecting enough attacked samples (which may be hidden among clean samples) to notice a statistically significant drop. Additionally, we have seen in Section~\ref{sec:stealthiness} that an attacker may optimise a suffix with less effect on response perplexity or quality, at the expense of attack performance. Both input-based and response-based detection defences additionally suffer from false positives, making such a filter costly in deployment.

Once detected, we need effective ways of defending against the attack. One approach is to change the target model, the drafter, or parameters specific to the drafting process such as the sampling strategy. Our experiments showed that the adversarial suffixes are robust to changes in sampling style and may transfer across drafters, but do not transfer well across target models. However, a change in target model may lead to a reduction in response quality or an increase in cost. Conversely, a change in drafting strategy may lead to less speculative acceleration, achieving the attacker's goal of slowing down speculative decoding. In the worst case, if the speed-up of speculative decoding is sufficiently degraded the victim may defensively switch to autoregressive decoding to avoid incurring the drafter overhead with no benefit.

\section{Conclusion and Future Work}

Operators of LLM applications commonly enable speculative decoding, as they trust it to reduce their costs without modifying model outputs. We have shown that speculative decoding is not reliable under adversarial input, and can be exploited to increase per-token costs, potentially beyond that of autoregressive decoding without speculative decoding enabled. Our attack, Speedbump, targets the expected number of draft tokens accepted per cycle. It degrades all seven speculative decoding methods we study, in the strongest case more than tripling the target-model forward passes needed per generated token. LLM application operators should therefore not assume that speculative savings hold for untrusted input, and should consider the suggested countermeasures for detecting and defending against potential Speculative Rejection Attacks.

Several extensions would bring our evaluation closer to realistic deployments. While we studied the key case of attacks under a batch size of 1, future work could investigate attacks under continuous batching, studying whether an attack could slow down co-batched requests from other users. We also used a single prompt template per task and a context containing only the attacker-controlled content. More realistic testing could evaluate the robustness of the adversarial suffixes to prompt template changes for the same task, as well as the effectiveness of the attack in scenarios where the attacker controls only part of the context.

Our suffixes were optimised against a single speculative decoding configuration. In practice, the attacker may not know the exact configuration of the victim, or may want their attack to slow down multiple configurations. Our acceptance proxies could firstly be extended to other verification rules, such as the typical acceptance used by Medusa and Hydra when sampling, against which both our attacks were weaker. Further work could examine whether an attacker can achieve improved transferability to unknown deployments by optimising suffixes against an ensemble of drafters, target models, and tasks. Alternatively, per-cycle acceptance revealed by timing side channels in streaming APIs \cite{carlini2024remote} may allow an attacker to optimise suffixes directly against a deployed system without surrogate models, motivating the need for defences that work against an attacker operating under such a threat model.

Since total cost scales with both output length and cost per token, future work could investigate whether SRAs can be combined with generation-prolonging attacks for a multiplicative effect.

Finally, further work could study the detection and mitigation of SRAs more rigorously, including whether adaptive attacks with fluency constraints can evade perplexity filtering. Defences such as adaptively switching or disabling speculative decoding also warrant evaluation, including the cost of such a defence in lost speed-up.

% 12 pages of body text (above)

\clearpage

\section*{LLM Usage Considerations}

LLMs were used for editorial purposes in this manuscript, and all outputs were inspected by the authors to ensure accuracy and originality. We used Claude for editing assistance and Claude Code as a coding assistant. All code was reviewed by the authors.
LLMs are the object of study in this work, so their use is unavoidable. We use only open-weight models (Vicuna-7B and Vicuna-13B v1.3), their publicly released drafters, and public datasets. No data was collected and no model was trained. We chose Vicuna as the only target family with released drafters for every method studied, and used the 7B and 13B models rather than 33B to reduce compute.
To limit the footprint, each suffix is optimised once per task and reused across inputs and decoding strategies.

% Optional
\section*{Ethical Considerations}

This work presents an attack that raises the inference cost of LLM applications that use speculative decoding, and therefore carries dual-use risk. We weighed this risk using the principles of the Menlo Report. Under Beneficence, we judge the benefits of disclosure to outweigh the harms. The attack surface follows directly from how speculative decoding works, and concurrent work has identified it independently. Service providers are better served by knowing that the speed-up can be degraded by adversarial content, so that they do not rely on it for capacity planning, and can monitor acceptance rates and adopt the mitigations we discuss. The potential harm is also bounded: an SRA can at most remove the speculative speed-up, leaving generation only slightly slower than autoregressive decoding, and it neither leaks user data nor produces harmful content. Under Respect for Persons and Respect for Law and Public Interest, all experiments were run on our own hardware with open-weight models and public datasets. No deployed system or third-party service was attacked, and the work involved no human subjects or personal data. We release our code for transparency and to support research on defences.

\bibliographystyle{IEEEtran}
\bibliography{references}

\appendices

\section{Greedy Results on Vicuna-7B}
\label{appendix:vicuna-7b}

Table \ref{tab:main_results_greedy_7b} contains the main metrics for suffixes optimised against Vicuna-7B for greedy decoding. Note that no pretrained head exists for EAGLE-3 on Vicuna-7B. The overall trends are similar to those of Vicuna-13B.

\begin{table*}[t]
\centering
\footnotesize
\caption{Attack results on Vicuna-7B across speculative decoding methods and datasets for greedy decoding. Each entry is formatted as clean/attacked. The final row reports the average absolute reduction marked by ↓, with the average relative reduction shown in parentheses.}
\label{tab:main_results_greedy_7b}
\setlength{\tabcolsep}{3pt}
\resizebox{\textwidth}{!}{%
\begin{tabular}{llccccccccc}
\toprule
& & \multicolumn{3}{c}{GSM8K} & \multicolumn{3}{c}{HumanEval} & \multicolumn{3}{c}{CNN/DM} \\
\cmidrule(lr){3-5} \cmidrule(lr){6-8} \cmidrule(lr){9-11}
Method & Attack & Speed-up$\downarrow$ & MAL$\downarrow$ & Acc. & Speed-up$\downarrow$ & MAL$\downarrow$ & Pass@1 & Speed-up$\downarrow$ & MAL$\downarrow$ & Rouge-L \\
\midrule
\multirow{2}{*}{SPS} & Speedbump-P & 1.51$\times$/1.26$\times$ & 2.47/2.44 & 13.7/3.1 & 1.61$\times$/1.26$\times$ & 2.62/2.10 & 14.7/2.0 & 1.58$\times$/1.09$\times$ & 2.49/1.77 & 23.1/16.8 \\
 & Speedbump-D & 1.51$\times$/1.31$\times$ & 2.47/1.92 & 13.7/11.1 & 1.61$\times$/1.33$\times$ & 2.62/2.06 & 14.7/12.1 & 1.58$\times$/1.51$\times$ & 2.49/2.43 & 23.1/23.3 \\
\addlinespace
\multirow{2}{*}{Medusa} & Speedbump-P & 1.95$\times$/1.25$\times$ & 2.59/1.67 & 13.9/4.0 & 2.13$\times$/1.99$\times$ & 2.79/2.67 & 14.7/11.5 & 1.38$\times$/1.20$\times$ & 2.03/1.62 & 23.0/12.2 \\
 & Speedbump-D & 1.95$\times$/1.72$\times$ & 2.59/2.13 & 13.9/7.6 & 2.13$\times$/1.94$\times$ & 2.79/2.62 & 14.7/9.2 & 1.38$\times$/1.39$\times$ & 2.03/1.86 & 23.0/19.7 \\
\addlinespace
\multirow{2}{*}{Hydra} & Speedbump-P & 2.30$\times$/1.63$\times$ & 3.70/2.94 & 13.9/5.2 & 2.66$\times$/2.43$\times$ & 3.92/3.64 & 14.7/12.6 & 1.77$\times$/1.29$\times$ & 2.77/1.89 & 23.0/12.5 \\
 & Speedbump-D & 2.30$\times$/1.73$\times$ & 3.70/2.76 & 13.9/5.6 & 2.66$\times$/2.42$\times$ & 3.92/3.70 & 14.7/8.0 & 1.77$\times$/1.61$\times$ & 2.77/2.63 & 23.0/21.2 \\
\addlinespace
\multirow{2}{*}{EAGLE} & Speedbump-P & 2.39$\times$/2.08$\times$ & 4.04/3.69 & 13.8/7.5 & 2.75$\times$/2.49$\times$ & 4.22/4.04 & 14.7/13.2 & 1.98$\times$/1.55$\times$ & 3.29/2.36 & 23.1/14.0 \\
 & Speedbump-D & 2.39$\times$/1.54$\times$ & 4.04/2.61 & 13.8/6.9 & 2.75$\times$/2.60$\times$ & 4.22/4.19 & 14.7/7.8 & 1.98$\times$/1.95$\times$ & 3.29/3.24 & 23.1/22.4 \\
\addlinespace
\multirow{2}{*}{EAGLE-2} & Speedbump-P & 2.96$\times$/2.16$\times$ & 5.02/4.05 & 13.7/6.6 & 3.14$\times$/2.84$\times$ & 5.36/5.18 & 14.7/11.5 & 2.07$\times$/1.47$\times$ & 3.96/2.29 & 23.0/8.1 \\
 & Speedbump-D & 2.96$\times$/1.78$\times$ & 5.02/3.36 & 13.7/5.6 & 3.14$\times$/3.02$\times$ & 5.36/5.25 & 14.7/2.3 & 2.07$\times$/2.03$\times$ & 3.96/3.52 & 23.0/20.6 \\
\addlinespace
PLD & Speedbump-P & 1.59$\times$/1.48$\times$ & 1.85/1.81 & 14.0/9.8 & 1.70$\times$/1.33$\times$ & 1.82/1.52 & 14.7/2.6 & 2.05$\times$/1.21$\times$ & 2.24/1.36 & 23.2/14.8 \\
\midrule
\multirow{2}{*}{Avg. drop} & Speedbump-P & {$\downarrow$0.47$\times$}\,{\scriptsize(21.4\%)} & {$\downarrow$0.51}\,{\scriptsize(14.6\%)} & {$\downarrow$7.8}\,{\scriptsize(56.4\%)} & {$\downarrow$0.27$\times$}\,{\scriptsize(13.0\%)} & {$\downarrow$0.26}\,{\scriptsize(9.2\%)} & {$\downarrow$5.8}\,{\scriptsize(39.5\%)} & {$\downarrow$0.50$\times$}\,{\scriptsize(27.1\%)} & {$\downarrow$0.92}\,{\scriptsize(31.8\%)} & {$\downarrow$10.0}\,{\scriptsize(43.4\%)} \\
 & Speedbump-D & {$\downarrow$0.61$\times$}\,{\scriptsize(25.1\%)} & {$\downarrow$1.01}\,{\scriptsize(26.8\%)} & {$\downarrow$6.4}\,{\scriptsize(46.6\%)} & {$\downarrow$0.20$\times$}\,{\scriptsize(8.9\%)} & {$\downarrow$0.22}\,{\scriptsize(7.2\%)} & {$\downarrow$6.8}\,{\scriptsize(46.4\%)} & {$\downarrow$0.06$\times$}\,{\scriptsize(3.2\%)} & {$\downarrow$0.17}\,{\scriptsize(5.7\%)} & {$\downarrow$1.6}\,{\scriptsize(7.0\%)} \\
\bottomrule
\end{tabular}}
\end{table*}

\section{Per-Seed Results for Attacking Greedy Decoding on Vicuna-13B}
\label{appendix:per_seed_greedy}

Table \ref{tab:per_seed_greedy_13b} contains the per-seed results for suffixes optimised against Vicuna-13B for greedy decoding. Note that while we recorded the MAL of the random suffixes, we didn't perform the paired timing so the random suffixes have no speedup metrics reported.

\begin{table*}[t]
\centering
\footnotesize
\caption{Per-seed attack results on Vicuna-13B across speculative decoding methods and datasets for greedy decoding. We report the metrics for the clean prompt, random suffix (averaged over three seeds), and the per-seed attack results.}
\label{tab:per_seed_greedy_13b}
\setlength{\tabcolsep}{4pt}
\begin{tabular}{llccccccccc}
\toprule
& & \multicolumn{3}{c}{GSM8K} & \multicolumn{3}{c}{HumanEval} & \multicolumn{3}{c}{CNN/DM} \\
\cmidrule(lr){3-5} \cmidrule(lr){6-8} \cmidrule(lr){9-11}
Method & Suffix & Speed-up$\downarrow$ & MAL$\downarrow$ & Acc. & Speed-up$\downarrow$ & MAL$\downarrow$ & Pass@1 & Speed-up$\downarrow$ & MAL$\downarrow$ & RL \\
\midrule
\multirow{8}{*}{SPS} & Clean & 1.54$\times$ & 2.38 & 26.5 & 2.10$\times$ & 2.75 & 14.7 & 1.57$\times$ & 2.44 & 24.3 \\
 & Random & -- & 2.37 & 23.9 & -- & 2.70 & 15.5 & -- & 2.43 & 24.1 \\
 & Speedbump-P (1) & 1.17$\times$ & 1.84 & 8.1 & 1.40$\times$ & 1.88 & 7.8 & 1.19$\times$ & 1.81 & 18.5 \\
 & Speedbump-P (2) & 1.47$\times$ & 2.24 & 19.3 & 1.63$\times$ & 2.38 & 12.1 & \textcolor{red}{0.98}$\times$ & 1.35 & 7.2 \\
 & Speedbump-P (3) & \textcolor{red}{0.96}$\times$ & 1.55 & 6.7 & 1.82$\times$ & 2.57 & 13.8 & 1.20$\times$ & 1.79 & 16.3 \\
 & Speedbump-D (1) & 1.33$\times$ & 2.00 & 19.7 & 1.50$\times$ & 2.18 & 12.9 & 1.45$\times$ & 2.23 & 23.9 \\
 & Speedbump-D (2) & 1.31$\times$ & 2.05 & 21.8 & 1.44$\times$ & 1.65 & 10.3 & 1.70$\times$ & 2.32 & 23.9 \\
 & Speedbump-D (3) & 1.39$\times$ & 2.07 & 24.9 & 1.58$\times$ & 1.83 & 12.1 & 1.54$\times$ & 2.30 & 24.1 \\
\midrule
\multirow{8}{*}{Medusa} & Clean & 2.02$\times$ & 2.65 & 26.5 & 2.35$\times$ & 2.70 & 14.7 & 1.52$\times$ & 2.11 & 24.2 \\
 & Random & -- & 2.66 & 23.8 & -- & 2.63 & 15.5 & -- & 2.11 & 24.1 \\
 & Speedbump-P (1) & 1.24$\times$ & 1.61 & 11.8 & 1.93$\times$ & 2.64 & 14.7 & 1.03$\times$ & 1.32 & 8.4 \\
 & Speedbump-P (2) & 1.57$\times$ & 1.98 & 17.1 & 2.00$\times$ & 2.59 & 15.5 & 1.01$\times$ & 1.27 & 7.1 \\
 & Speedbump-P (3) & 1.40$\times$ & 1.74 & 16.1 & 2.02$\times$ & 2.61 & 18.1 & 1.10$\times$ & 1.42 & 10.3 \\
 & Speedbump-D (1) & 1.41$\times$ & 2.01 & 19.1 & 1.93$\times$ & 2.19 & 15.5 & 1.51$\times$ & 1.94 & 22.8 \\
 & Speedbump-D (2) & 1.43$\times$ & 1.80 & 6.6 & 2.18$\times$ & 2.70 & 12.9 & 1.56$\times$ & 2.09 & 24.4 \\
 & Speedbump-D (3) & 1.45$\times$ & 1.90 & 13.3 & 2.05$\times$ & 2.71 & 16.4 & 1.13$\times$ & 1.35 & 12.4 \\
\midrule
\multirow{8}{*}{Hydra} & Clean & 2.50$\times$ & 3.74 & 26.5 & 2.58$\times$ & 3.86 & 14.7 & 1.92$\times$ & 2.88 & 24.2 \\
 & Random & -- & 3.73 & 23.9 & -- & 3.65 & 15.2 & -- & 2.88 & 24.2 \\
 & Speedbump-P (1) & 2.23$\times$ & 3.33 & 24.9 & 2.39$\times$ & 3.62 & 16.4 & 1.15$\times$ & 1.40 & 10.9 \\
 & Speedbump-P (2) & 2.14$\times$ & 3.37 & 18.0 & 2.41$\times$ & 3.47 & 14.7 & 1.56$\times$ & 2.23 & 17.6 \\
 & Speedbump-P (3) & 2.21$\times$ & 3.23 & 23.2 & 2.34$\times$ & 3.46 & 13.8 & \textcolor{red}{0.99}$\times$ & 1.43 & 7.7 \\
 & Speedbump-D (1) & 1.71$\times$ & 2.33 & 11.7 & 2.44$\times$ & 3.65 & 16.4 & 1.92$\times$ & 2.85 & 24.1 \\
 & Speedbump-D (2) & 1.59$\times$ & 2.29 & 17.6 & 2.41$\times$ & 3.61 & 15.5 & 1.80$\times$ & 2.79 & 24.0 \\
 & Speedbump-D (3) & 1.70$\times$ & 2.54 & 10.2 & 2.55$\times$ & 3.67 & 15.5 & 1.17$\times$ & 1.71 & 7.2 \\
\midrule
\multirow{8}{*}{EAGLE} & Clean & 2.62$\times$ & 4.03 & 26.8 & 2.76$\times$ & 4.27 & 14.7 & 2.13$\times$ & 3.39 & 24.2 \\
 & Random & -- & 4.02 & 23.6 & -- & 4.17 & 15.5 & -- & 3.38 & 24.1 \\
 & Speedbump-P (1) & 2.29$\times$ & 3.64 & 21.8 & 2.60$\times$ & 4.04 & 12.1 & 1.47$\times$ & 1.89 & 12.9 \\
 & Speedbump-P (2) & 1.67$\times$ & 2.56 & 14.2 & 2.58$\times$ & 4.06 & 15.5 & 1.22$\times$ & 1.78 & 7.6 \\
 & Speedbump-P (3) & 2.23$\times$ & 3.63 & 17.7 & 2.54$\times$ & 4.01 & 18.1 & 1.50$\times$ & 2.08 & 12.0 \\
 & Speedbump-D (1) & 1.75$\times$ & 2.93 & 9.0 & 2.67$\times$ & 4.04 & 17.2 & 1.88$\times$ & 2.83 & 18.3 \\
 & Speedbump-D (2) & 2.33$\times$ & 3.77 & 18.7 & 2.65$\times$ & 4.09 & 18.1 & 1.99$\times$ & 3.33 & 24.6 \\
 & Speedbump-D (3) & 2.45$\times$ & 3.88 & 22.0 & 2.70$\times$ & 4.11 & 13.8 & 1.99$\times$ & 3.02 & 19.8 \\
\midrule
\multirow{8}{*}{EAGLE-2} & Clean & 3.06$\times$ & 4.97 & 26.5 & 3.11$\times$ & 5.28 & 14.7 & 2.42$\times$ & 4.08 & 24.2 \\
 & Random & -- & 4.95 & 24.0 & -- & 5.14 & 15.5 & -- & 4.09 & 24.2 \\
 & Speedbump-P (1) & 2.11$\times$ & 3.33 & 15.8 & 2.93$\times$ & 4.83 & 15.5 & 2.42$\times$ & 4.07 & 25.2 \\
 & Speedbump-P (2) & 2.80$\times$ & 4.69 & 19.4 & 2.98$\times$ & 4.70 & 17.2 & 1.60$\times$ & 2.04 & 11.9 \\
 & Speedbump-P (3) & 2.48$\times$ & 4.26 & 3.6 & 2.94$\times$ & 4.86 & 13.8 & 1.27$\times$ & 1.78 & 9.7 \\
 & Speedbump-D (1) & 2.95$\times$ & 4.69 & 8.1 & 2.93$\times$ & 4.90 & 4.3 & 1.54$\times$ & 2.14 & 11.3 \\
 & Speedbump-D (2) & 2.57$\times$ & 4.01 & 20.3 & 3.16$\times$ & 5.16 & 10.3 & 2.36$\times$ & 4.01 & 23.4 \\
 & Speedbump-D (3) & 2.57$\times$ & 4.14 & 20.4 & 2.93$\times$ & 4.83 & 4.3 & 2.04$\times$ & 3.02 & 18.6 \\
\midrule
\multirow{8}{*}{EAGLE-3} & Clean & 3.44$\times$ & 5.93 & 26.5 & 3.60$\times$ & 6.36 & 14.7 & 3.40$\times$ & 5.68 & 24.2 \\
 & Random & -- & 5.89 & 23.9 & -- & 6.05 & 15.2 & -- & 5.68 & 24.2 \\
 & Speedbump-P (1) & 1.65$\times$ & 2.77 & 13.3 & 3.50$\times$ & 6.35 & 16.4 & 1.26$\times$ & 1.71 & 8.9 \\
 & Speedbump-P (2) & 3.28$\times$ & 5.41 & 20.0 & 3.51$\times$ & 5.92 & 16.4 & 1.54$\times$ & 1.81 & 10.5 \\
 & Speedbump-P (3) & 3.29$\times$ & 5.63 & 19.3 & 3.36$\times$ & 5.80 & 5.2 & 1.45$\times$ & 1.79 & 9.6 \\
 & Speedbump-D (1) & 2.13$\times$ & 5.10 & 10.8 & 3.53$\times$ & 6.05 & 14.7 & 3.14$\times$ & 5.63 & 24.1 \\
 & Speedbump-D (2) & 3.48$\times$ & 5.62 & 24.0 & 2.98$\times$ & 5.31 & 0.9 & 2.39$\times$ & 4.38 & 19.2 \\
 & Speedbump-D (3) & 3.17$\times$ & 5.44 & 16.8 & 3.54$\times$ & 6.23 & 13.8 & 1.25$\times$ & 1.86 & 9.1 \\
\midrule
\multirow{5}{*}{PLD} & Clean & 1.53$\times$ & 1.75 & 26.8 & 1.93$\times$ & 2.09 & 14.7 & 1.97$\times$ & 2.12 & 24.2 \\
 & Random & -- & 1.76 & 23.8 & -- & 2.07 & 15.2 & -- & 2.09 & 24.1 \\
 & Speedbump-P (1) & 1.55$\times$ & 1.70 & 20.0 & 1.46$\times$ & 1.60 & 8.6 & 1.13$\times$ & 1.28 & 17.5 \\
 & Speedbump-P (2) & 1.45$\times$ & 1.59 & 10.8 & 1.51$\times$ & 1.67 & 6.0 & 1.13$\times$ & 1.25 & 17.2 \\
 & Speedbump-P (3) & 1.60$\times$ & 1.67 & 19.0 & 1.48$\times$ & 1.66 & 12.1 & 1.09$\times$ & 1.28 & 18.3 \\
\bottomrule
\end{tabular}
\end{table*}

\section{Per-Seed Results for Attacking Sampling on Vicuna-13B}
\label{appendix:per_seed_sampling}

Table \ref{tab:per_seed_sampling_13b} contains the per-seed results for suffixes optimised against Vicuna-13B for sampling.

\begin{table*}[t]
\centering
\footnotesize
\caption{Per-seed attack results on Vicuna-13B across speculative decoding methods and datasets for sampling. We report the metrics for the clean prompt (averaged over three sampling seeds) and the per-seed attack results. We highlight cells where the speed-up is less than 1x in red.}
\label{tab:per_seed_sampling_13b}
\setlength{\tabcolsep}{4pt}
\begin{tabular}{llccccccccc}
\toprule
& & \multicolumn{3}{c}{GSM8K} & \multicolumn{3}{c}{HumanEval} & \multicolumn{3}{c}{CNN/DM} \\
\cmidrule(lr){3-5} \cmidrule(lr){6-8} \cmidrule(lr){9-11}
Method & Suffix & Speed-up$\downarrow$ & MAL$\downarrow$ & Acc. & Speed-up$\downarrow$ & MAL$\downarrow$ & Pass@1 & Speed-up$\downarrow$ & MAL$\downarrow$ & RL \\
\midrule
\multirow{7}{*}{SPS} & Clean & 1.28$\times$ & 1.99 & 20.4 & 1.38$\times$ & 2.11 & 10.4 & 1.32$\times$ & 2.12 & 23.3 \\
 & Speedbump-P (1) & \textcolor{red}{0.92$\times$} & 1.35 & 5.5 & 1.09$\times$ & 1.66 & 4.3 & 1.02$\times$ & 1.67 & 16.8 \\
 & Speedbump-P (2) & 1.16$\times$ & 1.96 & 7.3 & 1.33$\times$ & 2.03 & 6.9 & \textcolor{red}{0.79$\times$} & 1.24 & 6.9 \\
 & Speedbump-P (3) & \textcolor{red}{0.87$\times$} & 1.36 & 3.7 & 1.36$\times$ & 2.04 & 10.3 & \textcolor{red}{0.90$\times$} & 1.44 & 11.5 \\
 & Speedbump-D (1) & 1.10$\times$ & 1.65 & 11.5 & 1.14$\times$ & 1.70 & 12.9 & 1.13$\times$ & 1.81 & 22.5 \\
 & Speedbump-D (2) & \textcolor{red}{0.99$\times$} & 1.59 & 13.6 & 1.11$\times$ & 1.63 & 10.3 & 1.24$\times$ & 1.85 & 22.6 \\
 & Speedbump-D (3) & 1.03$\times$ & 1.64 & 12.7 & 1.14$\times$ & 1.73 & 12.1 & 1.17$\times$ & 1.84 & 23.4 \\
\midrule
\multirow{7}{*}{Medusa$^\dagger$} & Clean & 2.08$\times$ & 2.88 & 25.0 & 2.29$\times$ & 2.97 & 19.8 & 1.69$\times$ & 2.29 & 23.6 \\
 & Speedbump-P (1) & 1.42$\times$ & 1.78 & 13.7 & 2.18$\times$ & 2.91 & 12.9 & 1.18$\times$ & 1.50 & 8.4 \\
 & Speedbump-P (2) & 2.04$\times$ & 2.46 & 19.1 & 2.26$\times$ & 2.93 & 12.9 & 1.09$\times$ & 1.38 & 7.7 \\
 & Speedbump-P (3) & 1.84$\times$ & 2.24 & 16.3 & 2.36$\times$ & 2.92 & 14.7 & 1.39$\times$ & 1.61 & 11.8 \\
 & Speedbump-D (1) & 1.68$\times$ & 2.16 & 16.5 & 2.21$\times$ & 2.73 & 15.5 & 1.58$\times$ & 2.08 & 22.5 \\
 & Speedbump-D (2) & 1.65$\times$ & 2.13 & 8.0 & 2.24$\times$ & 2.97 & 16.4 & 1.60$\times$ & 2.26 & 23.9 \\
 & Speedbump-D (3) & 1.78$\times$ & 2.25 & 9.9 & 2.32$\times$ & 2.99 & 16.4 & 1.32$\times$ & 1.54 & 15.2 \\
\midrule
\multirow{7}{*}{Hydra$^\dagger$} & Clean & 2.74$\times$ & 4.02 & 24.0 & 2.76$\times$ & 4.12 & 12.9 & 2.16$\times$ & 3.20 & 22.8 \\
 & Speedbump-P (1) & 2.41$\times$ & 3.56 & 24.1 & 2.74$\times$ & 3.97 & 12.1 & 1.34$\times$ & 1.64 & 11.8 \\
 & Speedbump-P (2) & 2.47$\times$ & 3.77 & 18.3 & 2.87$\times$ & 3.82 & 18.1 & 2.01$\times$ & 2.86 & 18.3 \\
 & Speedbump-P (3) & 2.51$\times$ & 3.62 & 22.7 & 2.74$\times$ & 3.87 & 14.7 & 1.37$\times$ & 1.93 & 7.5 \\
 & Speedbump-D (1) & 2.32$\times$ & 3.46 & 9.4 & 2.82$\times$ & 4.06 & 13.8 & 2.07$\times$ & 3.13 & 23.0 \\
 & Speedbump-D (2) & 1.96$\times$ & 3.01 & 13.1 & 2.90$\times$ & 4.00 & 15.5 & 1.95$\times$ & 3.10 & 23.6 \\
 & Speedbump-D (3) & 2.27$\times$ & 3.54 & 7.2 & 2.72$\times$ & 4.01 & 17.2 & 1.35$\times$ & 1.93 & 7.2 \\
\midrule
\multirow{7}{*}{EAGLE} & Clean & 2.28$\times$ & 3.71 & 20.2 & 2.43$\times$ & 3.89 & 10.6 & 1.89$\times$ & 3.15 & 23.3 \\
 & Speedbump-P (1) & 1.88$\times$ & 3.37 & 13.3 & 2.42$\times$ & 3.78 & 8.6 & 1.03$\times$ & 1.64 & 9.1 \\
 & Speedbump-P (2) & 1.25$\times$ & 1.99 & 9.7 & 2.35$\times$ & 3.78 & 12.1 & \textcolor{red}{0.95$\times$} & 1.54 & 6.6 \\
 & Speedbump-P (3) & 2.04$\times$ & 3.36 & 9.2 & 2.49$\times$ & 3.81 & 12.9 & 1.27$\times$ & 1.79 & 10.6 \\
 & Speedbump-D (1) & 1.12$\times$ & 1.73 & 3.1 & 2.38$\times$ & 3.78 & 9.5 & 1.51$\times$ & 2.34 & 14.9 \\
 & Speedbump-D (2) & 1.89$\times$ & 3.27 & 10.2 & 2.38$\times$ & 3.75 & 11.2 & 1.87$\times$ & 3.08 & 21.2 \\
 & Speedbump-D (3) & 1.92$\times$ & 3.39 & 5.4 & 2.87$\times$ & 3.81 & 10.3 & 1.68$\times$ & 2.62 & 15.5 \\
\midrule
\multirow{7}{*}{EAGLE-2} & Clean & 2.74$\times$ & 4.54 & 20.5 & 2.89$\times$ & 4.83 & 12.1 & 2.15$\times$ & 3.84 & 23.3 \\
 & Speedbump-P (1) & 1.56$\times$ & 2.50 & 9.9 & 2.87$\times$ & 4.58 & 8.6 & 2.13$\times$ & 3.80 & 23.6 \\
 & Speedbump-P (2) & 2.14$\times$ & 3.98 & 4.2 & 2.97$\times$ & 4.57 & 8.6 & \textcolor{red}{0.95$\times$} & 1.45 & 5.7 \\
 & Speedbump-P (3) & 2.35$\times$ & 4.02 & 9.1 & 3.01$\times$ & 4.33 & 6.0 & 1.01$\times$ & 1.52 & 7.4 \\
 & Speedbump-D (1) & 2.20$\times$ & 4.00 & 3.0 & 2.55$\times$ & 4.08 & 2.6 & 1.14$\times$ & 1.89 & 8.4 \\
 & Speedbump-D (2) & 2.14$\times$ & 3.57 & 14.3 & 2.91$\times$ & 4.67 & 7.8 & 1.83$\times$ & 3.43 & 16.6 \\
 & Speedbump-D (3) & 1.76$\times$ & 2.86 & 13.8 & 2.68$\times$ & 4.40 & 4.3 & 1.29$\times$ & 1.99 & 10.3 \\
\midrule
\multirow{7}{*}{EAGLE-3} & Clean & 3.05$\times$ & 5.37 & 19.9 & 3.34$\times$ & 5.78 & 12.4 & 2.78$\times$ & 5.26 & 23.2 \\
 & Speedbump-P (1) & 1.11$\times$ & 1.71 & 4.8 & 3.48$\times$ & 5.73 & 16.4 & \textcolor{red}{0.84$\times$} & 1.46 & 6.4 \\
 & Speedbump-P (2) & 2.68$\times$ & 4.82 & 8.3 & 3.34$\times$ & 5.72 & 13.8 & \textcolor{red}{0.93$\times$} & 1.46 & 6.6 \\
 & Speedbump-P (3) & 2.74$\times$ & 4.95 & 8.7 & 3.08$\times$ & 5.02 & 6.9 & 1.11$\times$ & 1.55 & 7.5 \\
 & Speedbump-D (1) & 2.60$\times$ & 4.66 & 9.3 & 3.27$\times$ & 5.30 & 10.3 & 2.66$\times$ & 5.14 & 22.9 \\
 & Speedbump-D (2) & 2.60$\times$ & 4.77 & 11.0 & 2.67$\times$ & 4.59 & 0.0 & 1.68$\times$ & 3.02 & 15.2 \\
 & Speedbump-D (3) & 2.43$\times$ & 4.59 & 3.1 & 3.65$\times$ & 5.64 & 8.6 & 1.03$\times$ & 1.73 & 7.8 \\
\bottomrule
\end{tabular}
\end{table*}

\section{Dataset Prompts}
\label{appendix:prompts}

We illustrate the prompting template and the position of the injected adversarial suffix for the different tasks. Content taken from the dataset is denoted with curly braces \texttt{\{\}}, and the position of the injected adversarial suffix is denoted with \texttt{<<SUFFIX>>}. Recall that \texttt{<<SUFFIX>>} is constrained to start with a space, so does not fuse with the content it is placed after.

\subsection{GSM8K}

\begin{lstlisting}
Given the following problem, reason and give a final answer to the problem.
Problem: {question}<<SUFFIX>>
Your response should end with "The final answer is [answer]" where [answer] is the response to the problem.
\end{lstlisting}

\subsection{CNN/DM}

\begin{lstlisting}
Summarize the following article:

{article}<<SUFFIX>>

Summary:
\end{lstlisting}

\subsection{HumanEval}

\begin{lstlisting}
Please provide a self-contained Python script that solves the following problem in a markdown code block:
```python
{stub, with <<SUFFIX>> appended to the last content line of its docstring}
```
\end{lstlisting}

As an example:

\begin{lstlisting}
Please provide a self-contained Python script that solves the following problem in a markdown code block:
```python
from typing import List


def has_close_elements(numbers: List[float], threshold: float) -> bool:
    """ Check if in given list of numbers, are any two numbers closer to each other than
    given threshold.
    >>> has_close_elements([1.0, 2.0, 3.0], 0.5)
    False
    >>> has_close_elements([1.0, 2.8, 3.0, 4.0, 5.0, 2.0], 0.3)
    True<<SUFFIX>>
    """
```
\end{lstlisting}

\end{document}